\documentclass[lettersize, journal]{IEEEtran}
\IEEEoverridecommandlockouts
\usepackage{flushend}
\usepackage{algorithm}
\usepackage{algorithmic}
\usepackage{stfloats}
\usepackage{cite} 
\usepackage{amsthm}
\newtheorem{myRemark}{Remark}

\newtheorem{myDef}{Definition}

\usepackage{amsmath,amssymb,amsfonts,bm}
\allowdisplaybreaks
\usepackage{graphicx}
\usepackage{epstopdf}
\usepackage{subfigure}
\usepackage[caption=false, font=footnotesize, labelfont=sf, textfont=sf]{subfig}
\usepackage{textcomp}
\usepackage{stfloats}
\usepackage{url}
\usepackage{verbatim}
\usepackage{enumerate}
\usepackage{xcolor}
\def\BibTeX{{\rm B\kern-.05em{\sc i\kern-.025em b}\kern-.08em
    T\kern-.1667em\lower.7ex\hbox{E}\kern-.125emX}}
\usepackage{balance}

\begin{document}

\title{ RIS-aided Near-Field MIMO Communications: Codebook and Beam Training Design}
\author{Suyu Lv, ~\IEEEmembership{Graduate Student Member,~IEEE},
	Yuanwei Liu,~\IEEEmembership{Senior Member,~IEEE}, \\
        Xiaodong Xu,~\IEEEmembership{Senior Member,~IEEE},   
        Arumugam Nallanathan,~\IEEEmembership{Fellow,~IEEE},\\
         and A. Lee Swindlehurst, ~\IEEEmembership{Fellow,~IEEE}
\thanks{Suyu Lv is with the State Key Laboratory of Networking and Switching Technology, Beijing University of Posts and Telecommunications, Beijing, 100876, China (e-mail: lvsuyu@bupt.edu.cn).}
\thanks{Xiaodong Xu is with the State Key Laboratory of Networking and Switching Technology, Beijing University of Posts and Telecommunications, Beijing 100876, China, and also with the Department of Broadband Communication, Peng Cheng Laboratory, Shenzhen 518066, Guangdong, China (e-mail: xuxiaodong@bupt.edu.cn).}
\thanks{Yuanwei Liu and Arumugam Nallanathan are with the School of Electronic Engineering and Computer Science, Queen Mary University of London, London
E1 4NS, U.K. (e-mail: yuanwei.liu@qmul.ac.uk; a.nallanathan@qmul.ac.uk).}
\thanks{A. Lee Swindlehurst is with the Center for Pervasive Communications and Computing, Henry Samueli School of Engineering, University of California at Irvine, Irvine, CA 92697 USA (e-mail: swindle@uci.edu).}
}

\maketitle 

\begin{abstract}
Downlink reconfigurable intelligent surface (RIS)-assisted multi-input-multi-output (MIMO) systems are considered with far-field, near-field, and hybrid-far-near-field channels. 
According to the angular or distance information contained in the received signals, 1) a distance-based codebook is designed for near-field MIMO channels, based on which a hierarchical beam training scheme is proposed to reduce the training overhead; 2) a combined angular-distance codebook is designed for mixed-far-near-field MIMO channels, based on which a two-stage beam training scheme is proposed to achieve alignment in the angular and distance domains separately.
For maximizing the achievable rate while reducing the complexity, an alternating optimization algorithm is proposed to carry out the joint optimization iteratively.
Specifically, the RIS coefficient matrix is optimized through the beam training process, the optimal combining matrix is obtained from the closed-form solution for the mean square error (MSE) minimization problem, and the active beamforming matrix is optimized by exploiting the relationship between the achievable rate and MSE. 
Numerical results reveal that: 1) the proposed beam training schemes achieve near-optimal performance with a significantly decreased training overhead;  
2) compared to the angular-only far-field channel model, taking the additional distance information into consideration will effectively improve the achievable rate when carrying out beam design for near-field communications. 
\end{abstract}


\section{Introduction}

Millimeter wave (mmWave) communications (30 - 300 GHz) and Terahertz (THz) communications (0.1 - 10 THz) have been recognized as promising candidate technologies for future B5G/6G networks to realize the communication goals of ultra-wide bandwidth and ultra-high transmission rate  {\cite{MCOM.2018.1700928, MCOM.001.2000306, MCOMSTD.001.2000048}}. 
Although possessing extremely high bandwidth, mmWave and THz communication suffer from high propagation losses due to penetration, reflection and diffraction, resulting in a reduced link budget compared to that in the sub-6 GHz scenario {\cite{JSAC.2017.2720038}}.

To tackle the problems caused by high-frequency communications, reconfigurable intelligent surface (RIS) technology has received extensive research attention due to its distinctive capability of restructuring the wireless propagation environment and establishing additional reliable reflection-based links between the base station (BS) and the user {\cite{LWC.2020.2980225, JSAC.2021.3071836}}.
RISs consist of numerous passive controllable reflecting elements that can adjust the phase shift of incident signals with extremely low power consumption, and thus have been regarded as one of the main enabling solutions to achieve green communication and coverage extension for next-generation wireless systems  {\cite{JSAC.2020.3007211}}. 

The low-cost and low-complexity characteristics of RIS reflecting elements have also promoted the emergence of extremely large-scale RIS to realize higher array gain. 
However, with an increase in the number of RIS reflecting elements, the boundary separating the near-field communication (NFC) region and far-field communication (FFC) region becomes more distant from the RIS, making the near-field region non-negligible {\cite{TWC.2021.3126384, TWC.2022.3158894, TutorialReview}}.
In contrast to the {\textit{plane wave}} model in the far-field region, the electromagnetic structure in the near-field region is fundamentally different and  the {\textit{spherical wave}} model should be used {\cite{TWC.2021.3126384}}. 
The spherical-wave propagation model for NFC contains both angular and distance information, which allows the resulting beam patterns to focus on a specific point {\cite{TWC.2022.3158894}}. 
Thus, NFC can utilize the new dimension of distance to realize more precise signal enhancement, interference management and user localization for RIS-assisted wireless networks {\cite{TWC.2022.3218531, TAP.2022.3147533, LCOMM.2022.3215253, WCSP55476.2022.10039319}}. 

However, reliable phase shift design for RIS-assisted communication systems relies heavily on accurate channel state information (CSI), which is challenging to obtain due to the passive property of RISs.
In addition, the massive number of RIS elements in a multi-input-multi-output (MIMO) system will also lead to extremely large pilot overhead and high channel estimation complexity.
Compared to cascaded channel estimation based on least squares or minimum mean squared error, codebook-based beam training has been widely adopted for mmWave or THz communication systems {\cite{TCOMM.2017.2730878, TWC.2020.3019523, LWC.2022.3212344}}.
The beam training method can obtain CSI by estimating the physical directions of channel paths rather than the entire channel, thus realizing lower complexity.
Moreover, beam training can directly achieve reliable beamforming by selecting a phase-shift vector from among a set of pre-defined ``codewords" during the training procedure, and thus is considered to be an efficient approach to acquire CSI in RIS-assisted wireless communication systems {\cite{MWC.006.2100517, TCOMM.2023.3251374, TCOMM.2023.3278728}}.
Existing codebook design and beam training schemes cannot be directly applied to RIS-assisted NFC systems or hybrid FFC \& NFC systems.

\subsection{Prior Works}

\subsubsection{Channel modeling for NFC} 

Signals in the far field of an array lead to planar wavefronts that produce linear phase variations and negligible amplitude differences across the array. However, the situation is different for near-field signals.
In order to achieve better performance through reasonable resource allocation, accurate channel models are essential for NFC.
Hybrid-field channel modeling schemes were studied in {\cite{LCOMM2021_ChannelEstimation}} and {\cite{LCOMM2022_HybridField}}, where the far-field and near-field components are separately estimated. 
A polar-domain representation for near-field channel models was proposed in {\cite{TCOMM2022_ChannelEstimation}}, accounting for the information in both the angular and distance domains. 
Considering the different multi-path characteristics observed by antennas at different distances in large-scale MIMO NFC systems, the authors in {\cite{TAP.2022.3218759}} proposed a non-stationary distance modeling framework.

\subsubsection{Beam training schemes} 

The beam training process is implemented by searching for the optimal beam from multiple predefined directional beams (codewords).
For beam training, testing all codewords in the codebook during the training process is the most straightforward approach.
However, the codebook and the resulting training overhead can be extremely large for large-scale massive MIMO systems, especially when considering both angular and distance information in NFC. 
The authors in {\cite{TCOMM.2017.2730878}} proposed a new hierarchical codebook to achieve uniform beam alignment performance with low overhead for mmWave communication systems.
For mmWave and sub-THz systems with multiple users, {\cite{JSAC.2017.2720038}} and {\cite{TWC.2020.3019523}} proposed hierarchical beamforming training strategies to carry out simultaneous training for multiple users. 
Aiming at overcoming the resolution limit of directional angle estimation, {\cite{LWC.2022.3193877}} proposed a codebook-based beam search approach for mmWave massive MIMO transmissions.
However, these codebooks are designed for the far-field case.
Considering the spherical wavefront propagation model for NFC, {\cite{LWC.2022.3212344, TwoStage}} proposed a two-phase beam training method to sequentially perform the beam sweeping in the angular and distance domains.

\subsubsection{Beam training design for RISs} 

To implement beam training for RIS-based systems, one must first construct an appropriate RIS codebook based on information about the cascaded channel.
Then, based on the information obtained during the training phase, the RIS selects the optimal codeword from the predesigned codebook to perform passive beamforming.
For RISs with massive numbers of elements, highly efficient beam training schemes with good performance are required to reduce the training overhead. 
The authors in {\cite{TCOMM.2023.3251374}} first proposed a hierarchical codebook-generating method using pattern synthesis, then provided a hierarchical
beam training method using two multi-mainlobe codewords in each layer.
To train multiple users at the same time, a multi-lobe beam training mechanism was proposed in {\cite{TWC.2022.3174849}} based on the BS-RIS joint codebook, thus achieving reduced overhead. 
A multi-beam training method was proposed for a RIS-assisted multiuser system in {\cite{LWC.2020.3005980}}, by dividing the RIS into multiple sub-surfaces and designing multi-beam directions over time. 
Although  {\cite{TCOMM.2023.3278728, TGCN.2023.3259579, JCC.2022.06.015}} proposed beam training schemes for RIS-assisted NFC systems, only limited scenarios with single-antenna users were considered.

\subsection{Motivation and Contributions}

As the size of the RIS increases, the near-field region becomes non-negligible.
In NFC, CSI is not only related to the angle but also distance, thus the acquisition of CSI becomes very important for proper beam design.
The methods for obtaining CSI can be roughly classified into two categories: channel estimation and beam training {\cite{JCC.2022.06.015}}.
In this article, we consider the use of beam training to simultaneously estimate the channel parameters and perform passive beamforming {\cite{TWC.2017.2686402}}, which is simpler than performing channel estimation first and then separately determining the optimal beamforming. 

Although there are numerous beam training schemes for FFC, codebook and beam training design for NFC are still in their infancy, especially for RIS-assisted systems.
Moreover, the effect of different channel modeling methods for FFC and NFC has not been well investigated.
In order to fill this gap, in this article we focus on four different channel models for downlink MIMO RIS-aided systems. 
Furthermore, we design codebooks and beam training schemes for near-field and mixed-far-near-field channels, which are further used for RIS reflecting coefficient optimization.
The main contributions of our work are listed as follows:
\begin{enumerate} 
  \item We consider four different channel models for a downlink RIS-assisted MIMO system, namely, i) FF: far-field model for both BS-RIS and RIS-user links; ii) NF: near-field model for BS-RIS link, far-field model for RIS-user link; iii) FN: far-field model for BS-RIS link, near-field model for RIS-user link; iv) NN: near-field model for both BS-RIS and RIS-user links. 
   Here, the assumption of far-field propagation leads to a {\textit{planar-wave}} channel model, while near-field propagation leads to a {\textit{spherical-wave}} channel model.
  \item According to the angular or distance information contained in the received signals, we design i) a distance-based codebook for the NN channel model, based on which we propose a hierarchical beam training scheme to reduce the training overhead; ii) a combined angular-distance codebook for NF and FN channel models, based on which we propose a two-stage beam training scheme to separately achieve alignment in the angular and distance domains.
  \item To maximize the achievable rate, we propose an alternating optimization (AO) algorithm to carry out the joint optimization of the RIS coefficients and the transmit and receive beamformers in an iterative manner. Specifically, the RIS coefficient matrix is obtained by the proposed beam training schemes, the optimal combining matrix at the user is obtained from the closed-form solution for the mean square error (MSE) minimization problem, and the active beamforming matrix at the BS is optimized by exploiting the relationship between the achievable rate and the MSE. 
  \item We present numerical results revealing that: i) the proposed hierarchical and two-stage beam training approaches yield achievable rate performance similar to the exhaustive search (ES) method while significantly reducing the training overhead; ii) compared to angular-only far-field channel models, the achievable rate is improved by exploiting the additional distance information when performing the beam design for NFC.   
\end{enumerate}

The organization for the rest of this article is as follows.
In Section II, we describe the proposed RIS-assisted MIMO system and the four considered channel models.
In Section III, we design different codebooks for the different channel models based on the angular or distance information in the received signals.
In Section IV, we propose an AO algorithm to carry out the joint optimization in an iterative manner for maximizing the achievable rate. 
In particular, two beam training algorithms are proposed in Section IV-A based on the predesigned codebooks, the combining matrix optimization is given in Section IV-B, and the active beamforming matrix optimization is described in Section IV-C. 
Simulation results and conclusions are given in Section V and Section VI, respectively.

\section{System Model}

\begin{figure*}[t]
\begin{center}
\includegraphics[width=11.5cm]{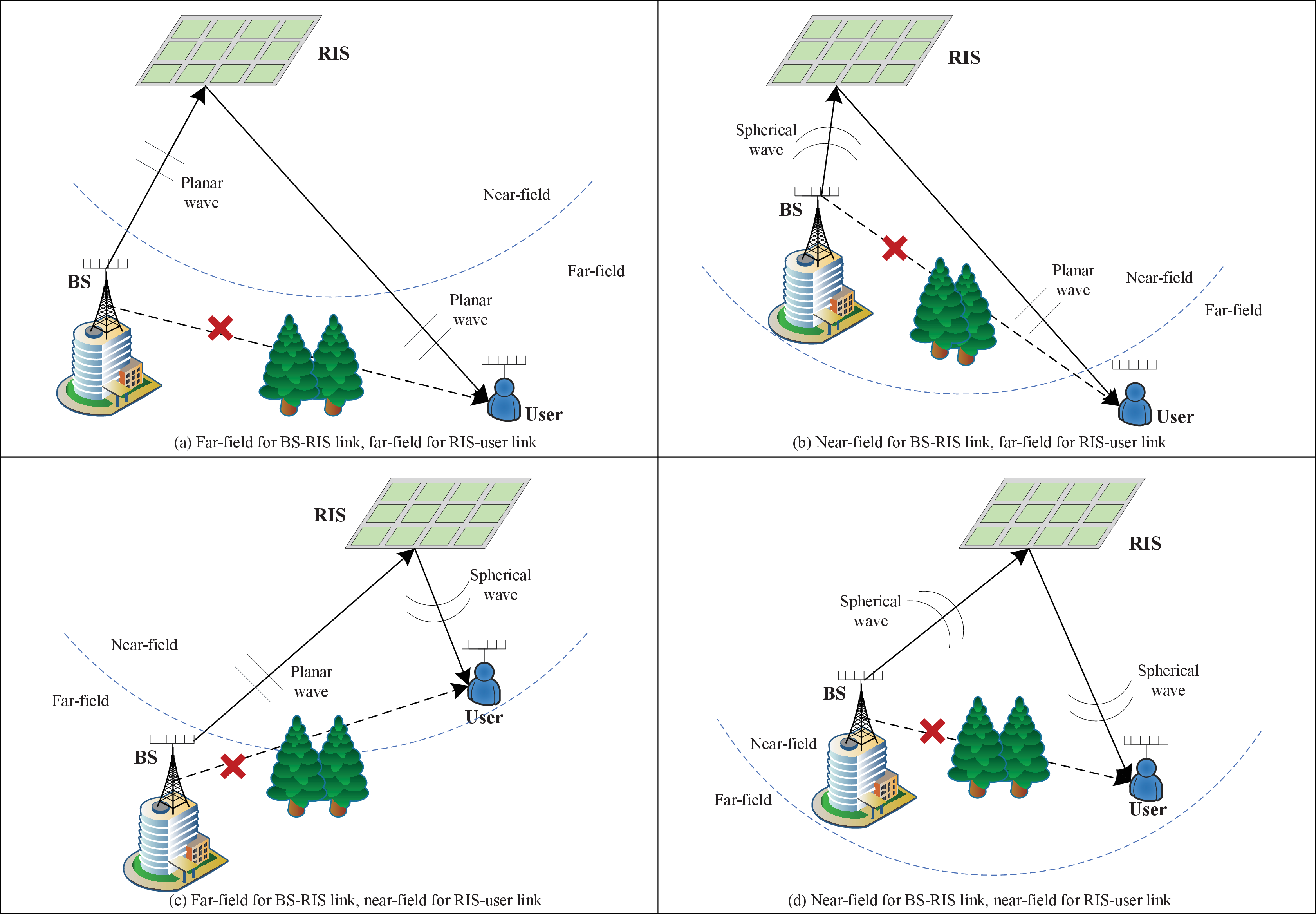}
\caption{Different channel models for RIS-assisted communication system.} 
\label{SystemModel_all}
\end{center}
\end{figure*}

We consider a RIS-assisted downlink communication scenario, where a single multi-antenna BS serves a multi-antenna user.
The direct path between the BS and the user is assumed to be blocked due to the existence of propagation obstacles.
The BS is equipped with an $N_{\rm{B}}$-element uniform linear array (ULA) of antennas, and the user is equipped with an $N_{\rm{U}}$-element ULA.
The RIS consists of a uniform planar array (UPA) with $M={M_x}{M_y}$ passive reflecting elements, where ${M_x}$ and ${M_y}$ denote the number of elements along the horizontal axis and vertical axis, respectively.

Four different channel models are considered for the RIS-assisted communication system, as shown in Fig. {\ref{SystemModel_all}}.
Specifically, these four channel models are respectively abbreviated as (a) FF: far-field model for both BS-RIS and RIS-user links; (b) NF: near-field model for BS-RIS link, far-field model for RIS-user link; (c) FN: far-field model for BS-RIS link, near-field model for RIS-user link; (d) NN: near-field model for both BS-RIS and RIS-user links.
We assume that the system operates at a frequency of $f_c$, so the wavelength of the signal is $\lambda_c = \frac{c}{f_c}$. 
Denote the array spacing as $d$, which we will assume here to be $d = \lambda_c/2$. 
Denote the array apertures of the BS, RIS and user as $D_B$, $D_R$ and $D_U$, respectively.
For the BS and the user with ULAs, the array apertures are calculated as $D_{\rm{B}} = \left(N_{\rm{B}}-1\right)d$ and $D_{\rm{U}} = \left(N_{\rm{U}}-1\right)d$,  respectively; for the RIS with a UPA, the array aperture is calculated as $D_{\rm{R}} = \sqrt{\left[\left(M_x-1\right)d\right]^2 + \left[\left(M_y-1\right)d\right]^2}$.
The comparison between the far-field region and near-field region in RIS-assisted systems is summarized in {\bf{TABLE} {\ref{tab1}}}, including the Rayleigh boundary when $d = \lambda_c/2$.

\begin{table*}[!t]
  \caption{Comparison between the far-field and near-field regions in RIS-assisted systems}
  \begin{center}
  \begin{tabular}{|c|c|c|}
  \hline
  \rule{0pt}{9pt}
   & BS-RIS channel & RIS-user channel \\
  \hline
  \rule{0pt}{9pt}
  Far-field & $r_{{\rm{B}}, {\rm{R}}}  > \frac{2 \left(D_{\rm{B}} + D_{\rm{R}}\right)^2 }{\lambda_c}$ & $ r_{{\rm{R}}, {\rm{U}}}  >  \frac{2 \left( D_{\rm{R}} + D_{\rm{U}} \right)^2 }{\lambda_c}$  \\
  \hline
  \rule{0pt}{9pt}
  Near-field & $ r_{{\rm{B}}, {\rm{R}}}  \le \frac{2 \left(D_{\rm{B}} + D_{\rm{R}}\right)^2 }{\lambda_c}$ & $ r_{{\rm{R}}, {\rm{U}}}  \le  \frac{2 \left( D_{\rm{R}} + D_{\rm{U}} \right)^2 }{\lambda_c}$  \\  
  \hline
  \rule{0pt}{9pt}
  $\begin{array}{cc}
        \rm{Boundary} \\
        \left(d = \lambda_c/2 \right) \\
  \end{array}$     &  $\frac{\left[ \left(N_{\rm{B}}-1\right) + \sqrt{\left(M_x-1\right)^2 + \left(M_y-1\right)^2} \right]^2{\lambda_c}}{2}$ & $\frac{\left[ \left(N_{\rm{U}}-1\right) + \sqrt{\left(M_x-1\right)^2 + \left(M_y-1\right)^2} \right]^2{\lambda_c}}{2}$ \\   
  \hline
   \end{tabular}
  \label{tab1}
  \end{center}
\end{table*}

\subsection{Channel Model}

In order to explore the differences among the four channel scenarios in terms of signal model, beam design, transmission performance, etc., we first provide mathematical descriptions of the models below.
More specifically, we will first present the far-field model for the BS-RIS and RIS-user channels, then give the near-field model for the BS-RIS and RIS-user channels, and finally obtain the cascaded channel models for the four channel scenarios mentioned above.

\subsubsection{Far-field channel model}

When the far-field channel model is considered, the channel between the BS and RIS is characterized by the Saleh-Valenzuela channel model \cite{TWC2014_SpatiallySparse}, given by 
\begin{equation}\label{ChannelGain_BR}
{{\bf{G}}_{{\rm{B}}, {\rm{R}}}^{\rm{far}}} = \sqrt {\frac{{{M}{N_{\rm{B}}}}}{L_{\rm{B}}}}   \sum\limits_{l = 0}^{L_{\rm{B}}-1} {{\beta _l}{\bf{a}}_{\rm{R}}\left( {\alpha _l^{\rm{A}}}, {\varphi _l^{\rm{A}}} \right) {{\bf{a}}_{\rm{B}}^H}\left( {\alpha _l^{\rm{D}}} \right) },
\end{equation}
where $L_{\rm{B}}$ is the number of paths between the BS and RIS, ${\beta _0}$ represents the complex gain of the line-of-sight (LoS) component, ${\beta _l} \left( 1\le l \le L_{\rm{B}} -1 \right)$ denotes the complex gain of the $l$-th non-line-of-sight (NLoS) path \cite{JSAC2014_mmWaveChannel},
${\alpha _l^{\rm{A}}}$ and $\varphi _l^{\rm{A}}$ denote the azimuth and elevation angle of arrival (AoA) associated with the RIS, ${\alpha _l^{\rm{D}}}$ denotes the angle of departure (AoD) associated with the BS, and ${\bf{a}}_{\rm{R}}\left( {\alpha _l^{\rm{A}}}, {\varphi _l^{\rm{A}}} \right)$ and ${{\bf{a}}_{\rm{B}}}\left( {\alpha _l^{\rm{D}}} \right)$ are the array response vectors associated with the RIS and the BS, respectively. 

For a UPA with $M_x$ and $M_y$ elements (${M_x}{M_y} = M$) on the horizontal and the vertical axes respectively, the RIS array response vector is given by
\begin{equation}\label{ArrayResponse_UPA}
\begin{array}{l}
{{\bf{a}}_{\rm{R}}}\left( {\alpha, \varphi } \right) = \frac{1}{{\sqrt {M} }}\Big[ 1, \cdots ,{e^{-j\frac{{2\pi d}}{\lambda_c }{\left( {{M_x} - 1} \right)\sin  \alpha  \sin \varphi  } }} \Big]^T   \\
\;\;\;\;\;\;\;\;\;\;\;\; \;\;\;\;\;\;\;\;\;\; \otimes \Big[ 1, \cdots ,e^{-j\frac{{2\pi d}}{\lambda_c } { \left( {{M_y} - 1} \right)\cos \varphi } } \Big]^T,
\end{array}
\end{equation}
where $0 \le x \le (M_x - 1)$ and $0 \le y \le (M_y - 1)$, and $\otimes$ is the Kronecker product. 
For the ULA with $N_{\rm{B}}$ elements at the BS, the array response vector is expressed as \cite{JSAC2016_mmWaveMIMO}
\begin{equation}\label{ArrayResponse_ULA}
{{\bf{a}}_{\rm{B}}}\left( \alpha  \right) = \frac{1}{{\sqrt {N_{\rm{B}}} }}{\left[ 1, \cdots ,{e^{-j\left( {N_{\rm{B}} - 1} \right)\frac{{2\pi d}}{\lambda_c }\sin \alpha }} \right]^T},
\end{equation}
where $0 \le n_b \le (N_{\rm{B}} - 1)$.
Since the response vector of a ULA is constant in the elevation domain, we omit the parameter ${\varphi }$ in the response vector of the BS in (\ref{ArrayResponse_ULA}).

Similarly, the channel between the RIS and user in the far-field region is given by
\begin{equation}
{\bf{G}}_{{\rm{R}}, {\rm{U}}}^{\rm{far}} =  \sqrt {\frac{{{M}{N_{\rm{U}}}}}{L_{\rm{U}}}}  \sum\limits_{l = 0}^{L_{\rm{U}}-1} {{\beta _l}{\bf{a}}_{\rm{U}}\left( {\alpha _l^{\rm{A}}} \right) {{\bf{a}}_{\rm{R}}^H}\left( {\alpha _l^{\rm{D}}}, {\varphi _l^{\rm{D}}} \right) },
\end{equation}
where $L_{\rm{U}}$ is the number of paths between the RIS and user, ${\alpha _l^{ {\rm{A}} }}$ denotes the azimuth AoA associated with the user, ${\alpha _l^{\rm{D}}}$ and ${\varphi _l^{\rm{D}}}$ denote the azimuth and elevation AoD associated with the RIS respectively, and the array response vectors ${\bf{a}}_{\rm{R}}\left( {\alpha _l^{\rm{D}}}, {\varphi _l^{\rm{D}}} \right)$ and ${\bf{a}}_{\rm{U}}\left( {\alpha _l^{ {\rm{A}} } } \right)$ can be generated similarly to (\ref{ArrayResponse_UPA}) and  (\ref{ArrayResponse_ULA}).

\begin{myRemark} \label{Remark_FarFieldRank}
Assume that the number of RIS elements is far larger than the number of BS and user antennas.
When the far-field channel model is considered (no matter whether the BS-RIS link or the RIS-user link follows the far-field model), the rank $r$ of the cascaded BS-RIS-user channel satisfies $r \le \min \left\{ L_{\rm{B}}, L_{\rm{U}} \right\}$, and we say that the channel has $r$ degrees of freedom (DoFs) {\cite{NFC_DOF}}. Thus, the dimension of the transmitted signal vector should satisfy $q \le r \le \min \left\{ L_{\rm{B}}, L_{\rm{U}} \right\}$.
When environmental scattering is absent and no NLoS paths are present, the rank of the cascaded channel is 1.
\end{myRemark}

\subsubsection{Near-field channel model}

For communications in the near-field domain, we consider a three-dimensional topology.
In FFC, the signal propagation paths from transmitters to receivers can be regarded as parallel to each other if they share the same AoA. 
However, in NFC, the angle distribution is also related to the distance between specific points of transmitters and receivers.
Assume that all the BS antennas are located on the $x$-axis, where the coordinate of the midpoint of the BS array is denoted as $\left( 0, 0, 0 \right) $. 
Thus, the coordinate of the $n_b$-th antenna at the BS can be denoted as 
\begin{equation}
{\bf{q}}_{\rm{B}}\left( n_b \right) = \left( {{\tilde n}_b}d, 0, 0 \right), 
\end{equation}
where ${{\tilde n}_b} = n_b - \frac{N_{\rm{B}} - 1}{2}$.

Similarly, assume that all of the user antennas are parallel to the $x$-axis, where the coordinate of the midpoint of the user array is $\left( x_{\rm{U}}, y_{\rm{U}}, z_{\rm{U}} \right) $. 
Thus, the location of $n_u$-th antenna at the user can be denoted as 
\begin{equation}
{\bf{q}}_{\rm{U}}\left( n_u \right) = \left( x_{\rm{U}} + {{\tilde n}_u}d, y_{\rm{U}}, z_{\rm{U}} \right), 
\end{equation}
where ${{\tilde n}_u} = n_u - \frac{N_{\rm{U}} - 1}{2}$.

Assume that all elements of the RIS are parallel to the XY-plane.
Denote the coordinate of the midpoint of the RIS as $\left( x_{\rm{R}}, y_{\rm{R}}, z_{\rm{R}} \right) $.
Thus, the coordinate of the $\left( m_x, m_y\right)$-th element of the RIS can be expressed as 
\begin{equation}
{\bf{q}}_{\rm{R}}\left( m_x, m_y \right) = \left( x_{\rm{R}}+{{\tilde m}_x}d, y_{\rm{R}}+{{\tilde m}_y}d, z_{\rm{R}} \right) ,
\end{equation} 
where ${{\tilde m}_x} = m_x - \frac{M_x - 1}{2}$ and ${{\tilde m}_y} = m_y - \frac{M_y - 1}{2}$.
Further, assume that the location and topology  of the RIS are known.

Based on the above geometry,  the LoS link for the near-field channel between BS and RIS can be modeled as {\cite{cui2021near}}
\begin{equation}
{{\bf{G}}_{{\rm{B}}, {\rm{R}}}^{\rm{near}}} = \left[ {\bf{g}}_{{\rm{B}}, {\rm{R}}}^1, \cdots, {\bf{g}}_{{\rm{B}}, {\rm{R}}}^{m_{xy}}, \cdots, {\bf{g}}_{{\rm{B}}, {\rm{R}}}^{M} \right]^T, 
\end{equation}
where ${\bf{g}}_{{\rm{B}}, {\rm{R}}}^{m_{xy}} = \Big[ \kappa_{m_{xy}, 1} e^{-j\frac{{2\pi }}{\lambda_c }d_{m_{xy},1}^{\rm{B}}}, \cdots, $ $ \kappa_{m_{xy}, N_{\rm{B}}} e^{-j\frac{{2\pi }}{\lambda_c }d_{m_{xy}, N_{\rm{B}}}^{\rm{B}}} \Big] ^T $ and $\kappa_{m_{xy}, n_b} \propto  \frac{1}{ d_{m_{xy}, n_b}^{\rm{B}}}$ denotes the free-space large-scale path loss between the $n_b$-th antenna of the BS and the $\left( m_x, m_y\right)$-th element of the RIS. 
The distance between the $n_b$-th antenna of the BS and the $\left( m_x, m_y\right)$-th element of the RIS is given by
\begin{equation}
d_{m_{xy}, n_b}^{\rm{B}} = \left \| {\bf{q}}_{\rm{B}}\left( n_b \right) -  {\bf{q}}_{\rm{R}}\left( m_x, m_y \right) \right \|_2, 
\end{equation}
which can be expanded as  
\begin{equation}
\begin{array}{l}
d_{m_{xy}, n_b}^{\rm{B}} = \sqrt{ \left( x_{\rm{R}}+{{\tilde m}_x}d - {{\tilde n}_b}d \right)^2 + \left( y_{\rm{R}}+{{\tilde m}_y}d \right)^2 + z_{\rm{R}}^2}  \\
\;\;\;\;\;\;\; = \sqrt{ {r_{m_{xy}}^{\rm{B}}}^2  + \left( {{\tilde n}_b}d \right)^2  - 2 {{\tilde n}_b}d r_{m_{xy}}^{\rm{B}} \sin  \alpha_{m_{xy}}^{\rm{B}}  \sin  \varphi_{m_{xy}}^{\rm{B}}  } \\
\;\;\;\;\;\;\; \overset{\left ( a \right ) }{\approx }  r_{m_{xy}}^{\rm{B}}  - {{\tilde n}_b}d \sin \alpha_{m_{xy}}^{\rm{B}} \sin \varphi_{m_{xy}}^{\rm{B}}  \\
\;\;\;\;\;\;\;\;\;\;\;  + \frac{\left( {{\tilde n}_b}d \right)^2 \left( 1- \sin^2 \alpha_{m_{xy}}^{\rm{B}} \sin^2 \varphi_{m_{xy}}^{\rm{B}} \right) }{2r_{m_{xy}}^{\rm{B}}},
\end{array} 
\end{equation}
with $r_{m_{xy}}^{\rm{B}}$ denoting the distance from $\left( 0, 0, 0 \right) $ and the $\left( m_x, m_y\right)$-th element of the RIS, $ \alpha_{m_{xy}}^{\rm{B}} $, $ \varphi_{m_{xy}}^{\rm{B}}$ denoting the corresponding azimuth and elevation angle, respectively,  and ${\left ( a \right ) }$ is obtained based on the first-order Taylor expansion $\sqrt{ 1 + x  } = 1 + \frac{1}{2}x - \frac{1}{8}x^2 + {\cal{O}} \left ( x^3 \right )$.

Similarly, the LoS near-field channel between the RIS and user can be modeled as
\begin{equation}
{{\bf{G}}_{{\rm{R}}, {\rm{U}}}^{\rm{near}}} = \left[ {\bf{g}}_{{\rm{R}}, {\rm{U}}}^1, \cdots, {\bf{g}}_{{\rm{R}}, {\rm{U}}}^{n_u}, \cdots, {\bf{g}}_{{\rm{R}}, {\rm{U}}}^{N_{\rm{U}}} \right]^T, 
\end{equation}
where ${\bf{g}}_{{\rm{R}}, {\rm{U}}}^{n_u} = \left[ \kappa_{n_u, 1} e^{-j\frac{{2\pi }}{\lambda_c }d_{n_u,1}^{\rm{U}}}, \cdots,  \kappa_{n_u, M} e^{-j\frac{{2\pi }}{\lambda_c }d_{n_u, M}^{\rm{U}}} \right] ^T $ and $\kappa_{n_u, m_{xy}} \propto \frac{1}{ d_{n_u, m_{xy}}^{\rm{U}}}$ denote the free-space large-scale path loss between the $n_u$-th antenna of the user and the $\left( m_x, m_y\right)$-th element of the RIS. 
The distance between the $n_u$-th antenna of the user and the $\left( m_x, m_y\right)$-th element of the RIS is given by
\begin{equation}
d_{n_u, m_{xy}}^{\rm{U}} = \left \| {\bf{q}}_{\rm{U}}\left( n_u \right) - {\bf{q}}_{\rm{R}}\left( m_x, m_y \right) \right \|_2,
\end{equation}
which can be approximated as  
\begin{equation}
\begin{array}{l}
d_{n_u, m_{xy}}^{\rm{U}} {\approx } \; r_{m_{xy}}^{\rm{U}}  - {{\tilde n}_u}d \sin\alpha_{m_{xy}}^{\rm{U}} \sin \varphi_{m_{xy}}^{\rm{U}} \\
\;\;\;\;\;\;\;\;\;\;\;\;\;\; + \frac{\left( {{\tilde n}_u}d \right)^2 \left( 1 - \sin^2\alpha_{m_{xy}}^{\rm{U}} \sin^2 \varphi_{m_{xy}}^{\rm{U}}\right) }{2r_{m_{xy}}^{\rm{U}}},
\end{array}
\end{equation}
with $r_{m_{xy}}^{\rm{U}}$ denoting the distance from the midpoint of the user array and the $\left( m_x, m_y\right)$-th element of the RIS, $\alpha_{m_{xy}}^{\rm{U}}$, $\varphi_{m_{xy}}^{\rm{U}}$ denoting the corresponding azimuth and elevation angle, respectively.

\begin{myRemark} \label{Remark_NearFieldRank}
When both the BS-RIS and RIS-user links are in the near-field region, the rank of the cascaded LoS channel is $r =\min \left\{ N_{\rm{B}}, N_{\rm{U}}, M \right\}$, which will be much larger than 1 in the case of large-scale arrays, even in the absence of NLoS paths.
Thus, one of the main advantages of NFC is that they possess higher DoFs without relying on abundant environmental scattering.
\end{myRemark}

\subsubsection{Cascaded channel model}

Denote the phase-shift matrix of the RIS as ${\bf{\Theta }} = {\rm diag} \big( \phi_1,  \cdots , $ $\phi_m , \cdots , \phi_M \big)$, with  $\phi_m = e^{-j\theta_m  }, \theta_m \in \left[ 0, 2\pi \right)$ denoting the phase-shift coefficient of the $m$-th RIS element.
Therefore, the cascaded channels of the four scenarios mentioned above can be expressed as
\begin{subequations}\label{CascadedChannel}
\begin{align}
1) \; & {\bf{H}}_{\rm{FF}} = {\bf{G}}_{{\rm{R}}, {\rm{U}}}^{\rm{far}} {\bf{\Theta }} {{\bf{G}}_{{\rm{B}}, {\rm{R}}}^{\rm{far}}}, \\
2) \; & {\bf{H}}_{\rm{NF}} = {\bf{G}}_{{\rm{R}}, {\rm{U}}}^{\rm{far}} {\bf{\Theta }} {{\bf{G}}_{{\rm{B}}, {\rm{R}}}^{\rm{near}}}, \\
3) \; & {\bf{H}}_{\rm{FN}} = {\bf{G}}_{{\rm{R}}, {\rm{U}}}^{\rm{near}} {\bf{\Theta }} {{\bf{G}}_{{\rm{B}}, {\rm{R}}}^{\rm{far}}}, \\
4) \; & {\bf{H}}_{\rm{NN}} = {\bf{G}}_{{\rm{R}}, {\rm{U}}}^{\rm{near}} {\bf{\Theta }} {{\bf{G}}_{{\rm{B}}, {\rm{R}}}^{\rm{near}}}.
\end{align}
\end{subequations}

\subsection{Signal Model}

We assume that fully-digital beamforming is adopted at the BS, and thus the BS transmitted signal is given by
\begin{equation}\label{DigitalBeamforming}
{\bf{s}} = {\bf{W}} {\bf{x}},
\end{equation}
where ${\bf{x}} \in {\mathbb{C}}^{q \times 1} $ is the symbol vector, satisfying ${\mathbb{E}}\left[ {\bf{x}} {\bf{x}}^H \right] = {\bf{I}}_{q}$, and
${\bf{W}}\in {\mathbb{C}} ^{N_{\rm{B}} \times q} $ is the beamforming matrix or precoder.

Assuming the receiver adopts a linear combiner, the received signal at the user is given by
\begin{equation}
{\bf{y}} = {\bf{U}}^H {\bf{H}} {\bf{W}} {\bf{x}} + {\bf{U}}^H{\bf{n}},
\end{equation}
where ${\bf{U}} \in {\mathbb{C}}^{N_{\rm{U}} \times q}$ is the combining matrix, ${\bf{H}} \in \left\{ {\bf{H}}_{\rm{FF}}, {\bf{H}}_{\rm{NF}}, {\bf{H}}_{\rm{FN}}, {\bf{H}}_{\rm{NN}} \right\}$ is the cascaded channel between the BS and the user based on the specific channel model, and ${\bf{n}} \in {\cal{CN}} \left( {\bf{0}}, \sigma^2{\bf{I}}_{N_{\rm{U}}} \right) $ is additive white Gaussian noise (AWGN).

The maximum achievable data rate (bits/s/Hz) for the user is given by
\begin{equation} \label{DataRate} 
R \left( {\bf{W}}, {\bf{\Theta }} \right) = \log_2 \left| {\bf{I}}_{N_{\rm{U}}} +  {\bf{H}} {\bf{W}} {\bf{W}}^H {\bf{H}}^H  \sigma^{-2} \right|,
\end{equation}
where $\left| {\bf{X}} \right|$ represents the determinant of the matrix ${\bf{X}}$. 

\begin{myRemark} \label{Remark_U}
Equation (\ref{DataRate}) characterizes the maximum achievable rate when the optimal combiner at the user, denoted as ${\bf{U}}_{\rm{opt}}$, is adopted.
Given $\left\{ {\bf{W}}, {\bf{\Theta }}\right\}$, closed-form solutions for ${\bf{U}}_{\rm{opt}}$ can be obtained, as detailed in the sequel.
\end{myRemark}

\section{RIS Codebook Design}

The precise design of the RIS coefficient matrix requires accurate channel information, which is difficult to obtain due to the passivity of RISs.
Therefore, in this article, we consider exploiting beam training, which can perform channel estimation and RIS coefficient optimization simultaneously. 
To begin with, in this section, we will first design codebooks for different channel models.
Based on the information contained in the received signals, angular-domain and distance-based codebooks are designed for FF and NN models respectively, while a combined angular-distance (AD) codebook is designed for NF and FN models.
Each column in the designed codebook corresponds to one candidate RIS passive beamforming vector.

\subsection{ Angular-domain Codebook for FF }

In this subsection, we consider an angular-domain codebook that is universal in the far-field domain. 
Generally, existing RIS codebooks for FF channel models aim at aligning the beam to the strongest propagation path to maximize the beamforming gain.
Ignoring weak paths, the signal at the user can be written as
\begin{equation}\label{y_FF}
\begin{array}{l}
{\bf{y}}_{\rm{FF}} = {\beta _0}'{\bf{U}}^H{{\bf{a}}_{\rm{U}}\left( {\alpha _0^{\rm{A}}} \right) {{\bf{a}}_{\rm{R}}^H}\left( {\alpha _0^{\rm{D}}}, {\varphi _0^{\rm{D}}} \right) } {\rm diag} \left({\bm{\phi }}\right)  \\
\;\;\;\; \;\;\;\;\; {{\bf{a}}_{\rm{R}}\left( {\alpha _0^{\rm{A}}}, {\varphi _0^{\rm{A}}} \right) {{\bf{a}}_{\rm{B}}^H}\left( {\alpha _0^{\rm{D}}} \right) } {\bf{W}} {\bf{x}} + {\bf{U}}^H {\bf{n}},
\end{array}
\end{equation}
where ${\beta _0}'$ denotes the complex gain of the cascaded LoS channel.
By substituting the array response vectors ${{\bf{a}}_{\rm{R}}}\left( {\alpha _0^{\rm{D}}}, {\varphi _0^{\rm{D}}} \right)$ and ${\bf{a}}_{\rm{R}}\left( {\alpha _0^{\rm{A}}}, {\varphi _0^{\rm{A}}} \right)$ into (\ref{y_FF}), the user's signal can be reformulated as 
\begin{equation}
\begin{array}{l}
{\bf{y}}_{\rm{FF}} = {\beta _0}'{\bf{U}}^H{\bf{a}}_{\rm{U}}\left( {\alpha _0^{\rm{A}}} \right) \sum\limits_{m = 1}^{M}  e^{ -j\left(\theta_m + \nu_{m}^{{\rm{A}}} - \nu_{m}^{{\rm{D}}} \right) }  {{\bf{a}}_{\rm{B}}^H}\left( {\alpha _0^{\rm{D}}} \right)  {\bf{W}} {\bf{x}} \\
\;\;\;\;\;\;\;\;\;\;\; + {\bf{U}}^H {\bf{n}},
\end{array}
\end{equation}
where $\nu_{m}^{{\rm{A}}} = \frac{{2\pi d }}{\lambda_c } \left[ \left( {{m_x} - 1} \right)\sin \alpha_0^{\rm{A}} \sin \varphi_0^{\rm{A}} + { \left( {{m_y} - 1} \right)\cos \varphi_0^{\rm{A}} }  \right] $ and  $\nu_{m}^{{\rm{D}}} = \frac{{2\pi d }}{\lambda_c } \big[ \left( {{m_x} - 1} \right)\sin \alpha_0^{\rm{D}} \sin \varphi_0^{\rm{D}} +  \left( {{m_y} - 1} \right)\cos \varphi_0^{\rm{D}}   \big] $.
Thus, the optimal RIS phase-shift coefficients are $\theta_m =  \nu_{m}^{{\rm{D}}} - \nu_{m}^{{\rm{A}}}$.
All candidate angle pairs on the beam grid should be considered when accurate channel state information is unavailable,  which is one of the primary principles for far-field codebook design. 

Assume that ${\bf{U}}$ and ${\bf{W}}$ have already been designed to align with the strongest path.
${\bf{U}}$ and ${\bf{W}}$ can be designed with the channel information obtained through beam training, and we will show how ${\bf{U}}$ and ${\bf{W}}$ can be obtained later.
Thus, the RIS codebook for the FF channel model can be designed as  {\cite{LWC.2020.3005980}} {\cite{JCC.2022.06.015}}
\begin{equation}\label{Codebook_FFC}
\begin{array}{l}
{\cal{F}}_{\rm{FF}}=\Big[ {\bf{b}}\left( \beta_1, \delta_1 \right), \cdots, {\bf{b}}\left( \beta_{m_x}, \delta_{m_y} \right), \cdots, {\bf{b}}\left( \beta_{M_x}, \delta_{M_y} \right) \Big]^*,
\end{array}
\end{equation}
where $\beta_{m_x} = \frac{2{m_x}-{M_x}-1}{M_x}$ with ${m_x} = 1, \cdots, {M_x}$,  $\delta_{m_y} = \frac{2{m_y}-{M_y}-1}{M_y}$ with ${m_y} = 1, \cdots, {M_y}$, and ${{\bf{b}}}\left( {\beta, \delta } \right)$ is the following far-field steering vector
\begin{equation}\label{steering_FFC}
\begin{array}{l}
{{\bf{b}}}\left( {\beta, \delta } \right) = \Big[ 1, \cdots ,{e^{-j\frac{{2\pi d}}{\lambda_c }{\left( {{M_x} - 1} \right) \beta  \delta} }} \Big]^T  \otimes  \\
\;\;\;\;\;\;\;\;\;\;\;\;\;\;\;\;\;\; {\Big[ 1, \cdots ,e^{-j\frac{{2\pi d}}{\lambda_c } { \left( {{M_y} - 1} \right) \delta } }} \Big]^T. 
\end{array}
\end{equation}
Each column of ${\cal{F}}_{\rm{FF}}$ is a candidate beam codeword for the RIS in the far-field domain. 
This angular-domain codebook design makes full use of the angular information of the far-field paths.

\subsection{ Distance-based Codebook Design for NN }

Angular-only far-field codebooks are not applicable to near-field channels, even if only one or the other of the BS-RIS and RIS-user links is in the near-field domain because both distance and angular information are embedded in the received signal. 
Therefore, novel codebooks are required to match the channels in the near-field region. 
When both the BS-RIS link and RIS-user link are in the near-field domain, the combined signal at the user is given by (\ref{y_NN}) at the bottom of the next page, 
where $\nu_{n_u, m, n_b} =  \frac{{2\pi }}{\lambda_c }\left( d_{n_u, m}^{\rm{U}} + d_{m, n_b}^{\rm{B}}\right)$, determined by the sum of the distances from the BS to RIS and from the RIS to the user.

\begin{figure*}[!b]
\hrulefill
\begin{equation}\label{y_NN}
\begin{array}{l}
{\bf{y}}_{{\rm{NN}}} =\left( \frac{ \lambda_c}{4\pi} \right)^2 {\bf{U}}^H
\begin{bmatrix}
\sum\limits_{m = 1}^{M} \frac{1}{ d_{1, m}^{\rm{U}}d_{m, 1}^{\rm{B}} } e^{ -j\left( \theta_m + \nu_{1, m, 1} \right) } & \cdots & \sum\limits_{m = 1}^{M} \frac{1}{ d_{1, m}^{\rm{U}}d_{m, N_{\rm{B}}}^{\rm{B}} } e^{ -j\left( \theta_m + \nu_{1, m, N_b} \right) }  \\
  \vdots &  & \vdots \\
\sum\limits_{m = 1}^{M} \frac{1}{ d_{n_u, m}^{\rm{U}}d_{m, 1}^{\rm{B}} } e^{ -j\left( \theta_m + \nu_{n_u, m, 1} \right) } & \cdots & \sum\limits_{m = 1}^{M} \frac{1}{ d_{n_u, m}^{\rm{U}}d_{m, N_{\rm{B}}}^{\rm{B}} } e^{ -j\left( \theta_m + \nu_{n_u, m, N_{\rm{B}}} \right) } \\
  \vdots &  & \vdots \\
\sum\limits_{m = 1}^{M} \frac{1}{ d_{N_{\rm{U}}, m}^{\rm{U}}d_{m, 1}^{\rm{B}} } e^{ -j\left( \theta_m + \nu_{N_{\rm{U}}, m, 1} \right) } & \cdots & \sum\limits_{m = 1}^{M} \frac{1}{ d_{N_{\rm{U}}, m}^{\rm{U}}d_{m, N_{\rm{B}}}^{\rm{B}} } e^{ -j\left( \theta_m + \nu_{N_{\rm{U}}, m, N_{\rm{B}}} \right) } \end{bmatrix} 
 {\bf{W}} {\bf{x}}  + {\bf{U}}^H{\bf{n}},
\end{array}
\end{equation}
\end{figure*}

Define the following array steering vector related to the distance 
\begin{equation}\label{steering_d}
\begin{array}{l}
{\bm{f}}\left( x, y, z \right) = \exp \left[ -j\frac{{2\pi }}{{\lambda_c }} \left( r_{1, 1}, \cdots, r_{m_x, m_y}, \cdots, r_{M_x, M_y}\right) \right]^T , 
\end{array}
\end{equation}
where 
$ \left( x, y, z \right)$ denotes the coordinate of the BS or the user, and $r_{m_x, m_y} = \left \| \left( x, y, z \right),  {\bf{q}}_{\rm{R}}\left( m_x, m_y \right) \right \|_2$.
Generally, the height of the BS or the user is constant.
Therefore, for the design of the near-field codebooks, we mainly consider the exploration of $x$ and $y$, and the $z$ component in ${\bm{f}}\left( x, y, z \right)$ will be omitted hereafter.

\begin{myDef}\label{Def_operation}
The $ \star $ operator is defined as follows for vectors of equal dimensions:
\begin{equation}\label{OperationStar}
\begin{array}{l}
\;\;\;\; \left[ {\bf{x}}_1, \cdots, {\bf{x}}_M \right] \star \left[ {\bf{y}}_1, \cdots,  {\bf{y}}_N \right]  \\
= \left[ {\bf{x}}_1\odot{\bf{y}}_1, \cdots, {\bf{x}}_1\odot{\bf{y}}_N, \cdots, {\bf{x}}_M\odot{\bf{y}}_N \right],
\end{array}
\end{equation}
where $\odot$ denotes the Hadamard product.
\end{myDef}
 
We use a pair of coordinates to represent the rectangular sampling range on the XY-plane as $\left( x^{k}_{\min}, y^{k}_{\min} \right)$ and $\left( x^{k}_{\max}, y^{k}_{\max} \right)$, where $x^{k}_{\min}$ ($y^{k}_{\min}$) and $x^{k}_{\max}$ ($y^{k}_{\max}$) are the minimum and the maximum $x$ ($y$)-coordinate of the sampling points respectively, and $k \in \left\{ {\rm{B}}, {\rm{U}}\right\}$ indicates the BS or user.
Denote the number of points sampled on the $x$-axis and $y$-axis as $S_x$ and $S_y$, respectively.
Taking the corresponding distance between the sampling points into account, the codebook at the RIS for the NN channel model is given by
\begin{equation}\label{Codebook_NN}
\begin{array}{l}
{\cal{F}}_{\rm{NN}} =  \Big[ {\bm{f}}\left( x_1^{\rm{U}}, y_1^{\rm{U}} \right), \cdots, {\bm{f}}\left( x_{s_x}^{\rm{U}}, y_{s_y}^{\rm{U}} \right), \cdots, {\bm{f}}\left( x_{S_x}^{\rm{U}}, y_{S_y}^{\rm{U}} \right) \Big]^*  \\ 
\;\;\;\; \;\;\;\; \;\; \star  \Big[ {\bm{f}}\left( x_1^{\rm{B}}, y_1^{\rm{B}} \right), \cdots, {\bm{f}}\left( x_{s_x}^{\rm{B}}, y_{s_y}^{\rm{B}} \right), \cdots, {\bm{f}}\left( x_{S_x}^{\rm{B}}, y_{S_y}^{\rm{B}} \right) \Big]^*,
\end{array}
\end{equation}
where $x_{s_x}^k$ and $y_{s_x}^k$ are given by
\begin{subequations}\label{RangeSampling}
\begin{equation}\label{RangeSampling_x}
x_{s_x}^{k} = x^{k}_{\min} + \left( s_x - \frac{1}{2} \right) \frac{{x^{k}_{\max} - x^{k}_{\min} }}{{S_x}}, \; s_x = 1,\cdots, S_x,
\end{equation}
\begin{equation}\label{RangeSampling_y}
y_{s_y}^{k} = y^{k}_{\min} + \left( s_y - \frac{1}{2} \right) \frac{{y^{k}_{\max} - y^{k}_{\min} }}{{S_y}}, \; s_y = 1,\cdots, S_y.
\end{equation}
\end{subequations}
The size of the codebook will increase with the number of sampling points in each direction, and the accuracy of the RIS beamforming will improve correspondingly.

\subsection{ Combined AD Codebook Design for Hybrid FFC \& NFC }

In this section, we consider hybrid cascaded channel models, where one side of the RIS is in the far-field and the other is in the near-field.
As such, the link on one side of RIS only needs to consider angular information, while the link on the other side needs to consider both angular and distance information.
Taking the FN channel model as an example, i.e., with the BS-RIS link in the far-field and the RIS-user link in the near-field, the received signal at the user is given by
\begin{equation}\label{received_FN}
\begin{array}{l}
{\bf{y}}_{{\rm{FN}}} = {\bf{U}}^H {{\bf{G}}_{{\rm{R}}, {\rm{U}}}^{\rm{near}}} {\rm diag} \left({\bm{\phi }}\right) {{\bf{G}}_{{\rm{B}}, {\rm{R}}}^{\rm{far}}} {\bf{W}} {\bf{x}} + {\bf{U}}^H {\bf{n}}\\
\;\;\; = \frac{{\beta _0} \lambda_c}{4\pi{\sqrt {M} }} {\bf{U}}^H
\bigg[ \sum\limits_{m = 1}^{M} \frac{1}{ d_{1, m}^{\rm{U}}} e^{ -j\left(\theta_m + \nu_{1, m}^{{\rm{A}}} \right) } 
 , \cdots,   \\
\;\;\;  \sum\limits_{m = 1}^{M} \frac{1}{ d_{N_{\rm{U}}, m}^{\rm{U}}} e^{-j\left(\theta_m + \nu_{N_{\rm{U}}, m}^{{\rm{A}}} \right)} \bigg]^T
{{\bf{a}}_{\rm{B}}^H}\left( {\alpha _0^{\rm{D}}} \right)  {\bf{W}} {\bf{x}}  + {\bf{U}}^H {\bf{n}}, 
\end{array}
\end{equation}
where
\begin{equation}
\begin{array}{l}
\nu_{n_u, m}^{\rm{A}} = \frac{{2\pi }}{\lambda_c } \bigg[ \left( {{m_x} - 1} \right)d\sin \alpha_0^{\rm{A}} \sin \varphi_0^{\rm{A}} \\
\;\;\;\;\;\;\;\;\;\;\;\;\; + { \left( {{m_y} - 1} \right)d\cos \varphi_0^{\rm{A}} }  + d_{n_u, m}^{\rm{U}} \bigg].
\end{array}
\end{equation}
Similar to the assumptions above for FF channels, 
we assume that the transmit or receive beam related to the far-field link has been designed to align with the steering vector of the strongest path.
That is to say, $ {\bf{W}} $ has been aligned to ${{\bf{a}}_{\rm{B}}^H}\left( {\alpha _0^{\rm{D}}} \right) $ when designing the RIS codebook.

Considering the array response in both the distance and angular domains, we define the array steering vector as 
\begin{equation}\label{steering_polar}
\begin{array}{l}
{{\bf{c}}}\left( {x, y, \beta, \delta } \right) = \Big[ {e^{-j\frac{{2\pi }}{\lambda_c }d_{1,1} }}, \cdots, {e^{-j\frac{{2\pi }}{\lambda_c }d_{M_x, M_y} }}  \Big]^T \odot   \\ 
\left\{ \Big[ 1, \cdots ,{e^{-j\frac{{2\pi d}}{\lambda_c }{\left( {{M_x} - 1} \right) \beta  \delta} }} \Big]^T \otimes {\Big[ 1, \cdots ,e^{-j\frac{{2\pi d}}{\lambda_c } { \left( {{M_y} - 1} \right) \delta } }} \Big]^T \right\}, 
\end{array}
\end{equation}
where $d_{m_x, m_y} = \left \| \left( x, y, z \right),  {\bf{q}}_{\rm{R}}\left( m_x, m_y \right) \right \|_2$ and $z$ is assumed to be fixed and known.

A codebook design for hybrid cascaded channel models should cover not only the angular region of all possible signals, but their distances as well. 
Therefore, based on the received signal model given in ({\ref{received_FN}}),  the codebook for hybrid cascaded channel models should be designed to include the following two components.

\subsubsection{Angular-domain sweeping, similar to FFC}
To align the phase on the FFC side, the full angular range will be covered in the predesigned codebook. 
Therefore, the first component of the RIS codebook for hybrid cascaded channel models is given by
\begin{equation}\label{Codebook_Hy1}
\begin{array}{l}
{\cal{F}}_{\rm{Hb}}^1=\Big[ {\bf{b}}\left( \beta_1, \delta_1 \right), \cdots, {\bf{b}}\left( \beta_{m_x}, \delta_{m_y} \right), \cdots, {\bf{b}}\left( \beta_{M_x}, \delta_{M_y} \right) \Big]^*,
\end{array}
\end{equation}
where $\beta_{m_x} = \frac{2{m_x}-{M_x}-1}{M_x}$ with ${m_x} = 1, \cdots, {M_x}$,  $\delta_{m_y} = \frac{2{m_y}-{M_y}-1}{M_y}$ with ${m_y} = 1, \cdots, {M_y}$.

\subsubsection{Distance-domain sampling, similar to NFC}
For each given angle pair $\beta$ and $\delta$, all possible sampling points in the feasible range should be included to align the phase on the NFC side.
Therefore, the second component of the RIS codebook for hybrid cascaded channel models is given by 
\begin{equation}\label{Codebook_Hy2}
\begin{array}{l}
{\cal{F}}_{\rm{Hb}}^{2, k} = \Big[ {\bm{f}}\left( x_1^k, y_1^k \right), \cdots, {\bm{f}}\left( x_{s_x}^k, y_{s_y}^k \right), \cdots, {\bm{f}}\left( x_{S_x}^k, y_{S_y}^k \right) \Big]^*,
\end{array}
\end{equation}
where $x_{s_x}^k$ and $y_{s_x}^k$, $k \in \left\{ {\rm{B}}, {\rm{U}}\right\}$, are defined as in (\ref{RangeSampling}).
In particular, $k={\rm{B}}$ for the NF case, and $k={\rm{U}}$ for the FN case.

Accordingly, the codebook at the RIS for hybrid cascaded channel models can be designed as
\begin{subequations}
\begin{equation}\label{Codebook_NF}
{\cal{F}}_{\rm{NF}} = {\cal{F}}_{\rm{Hb}}^1 \star {\cal{F}}_{\rm{Hb}}^{2, {\rm{B}}},
\end{equation}
\begin{equation}\label{Codebook_FN}
{\cal{F}}_{\rm{FN}} = {\cal{F}}_{\rm{Hb}}^1 \star {\cal{F}}_{\rm{Hb}}^{2, {\rm{U}}},
\end{equation}
\end{subequations}
where the definition of $\star$ is given in (\ref{OperationStar}).

\section{AO-based Algorithm for Maximizing Rate}

We aim to maximize the achievable rate by jointly optimizing the combining matrix $\bf{U}$ at the user, the digital beamforming matrix ${\bf{W}}$ at the BS, and the reflecting coefficient matrix ${\bf{\Theta }}$ at the RIS.
Suppose each passive beamforming vector at the RIS is selected from the predesigned codebooks.
Therefore, the mathematical form of the optimization problem can be given as
\begin{subequations}\label{P_re}
\begin{align}
({\rm{P}})\; & \max \limits_{ \bf{W}, {\bm{\phi }} } R \left( {\bf{W}},  {\rm diag} \left( {\bm{\phi }} \right) \right), \\
s.t. \; &  \left \| {\bf{W}} \right \|_F^2 \le P_{\max}, \label{P_max}  \\
& {\bm{\phi }} \in  {\cal{F}}, \label{CodebookConstraint}
\end{align} 
\end{subequations}
where ${\cal{F}} \in\left\{ {\cal{F}}_{\rm{FF}}, {\cal{F}}_{\rm{NF}}, {\cal{F}}_{\rm{FN}}, {\cal{F}}_{\rm{NN}}\right\} $ in (\ref{CodebookConstraint}) is the predesigned codebook based on the specific channel model. 

To handle the highly-coupled non-convex problem in (\ref{P_re}), we propose an alternating optimization algorithm in this section, where three sub-problems will be solved in an iterative manner.
First, based on the predesigned codebooks, different beam training schemes will be proposed to perform the channel estimation and the RIS optimization for the specific channel models.
Second, a closed-form solution for the receive combining matrix at the user is obtained.
Finally, the active beamforming problem is solved by exploiting the relationship between the achievable rate and the mean square error using convex optimization.
Details for solving these sub-problems are given in the following subsections.

\subsection{Beam Training Design for RIS} 

\subsubsection{ Angular-domain Beam Sweeping for FF }   \label{AngularSweeping} 

The most direct method to perform beam training for this case is sweeping through all possible angles if the resulting training overhead is acceptable.
The size of the FF codebook in (\ref{Codebook_FFC}) depends on the number of RIS reflecting elements.
Based on the given angle pair $\left \langle {\beta_{m_x}, \delta_{m_y} } \right \rangle$, the corresponding codeword is ${{\bf{b}}}\left( {\beta_{m_x}, \delta_{m_y}} \right)$, and the achievable rate can be calculated as
\begin{equation}
R \left( {\bf{W}}, {\rm diag} \left( {{\bf{b}}}\left( {\beta_{m_x}, \delta_{m_y}} \right) \right) \right),
\end{equation}
with given beamforming matrix ${\bf{W}}$.
The optimal codeword is selected as the one that yields the maximum rate after sweeping through all candidate codewords in (\ref{Codebook_FFC}).

\subsubsection{ Hierarchical Beam Training Scheme for NN }   \label{HierarchicalTraining} 

From (\ref{Codebook_NN}) and (\ref{RangeSampling}), we see that the NN codebook size is determined by the number of sampled points on the $x$-axis and $y$-axis. 
The beam training overhead can be reduced by reducing the codebook size (i.e., increasing the step size between the codebook samples), but this will reduce the performance of the beam training.
In order to solve this problem, we design a hierarchical beam training approach that consists of several different levels of sub-codebooks. 
These sub-codebook levels are determined using different sampling ranges and sampling step sizes.

\begin{figure}[t]
\begin{center}
\includegraphics[width=8.5cm]{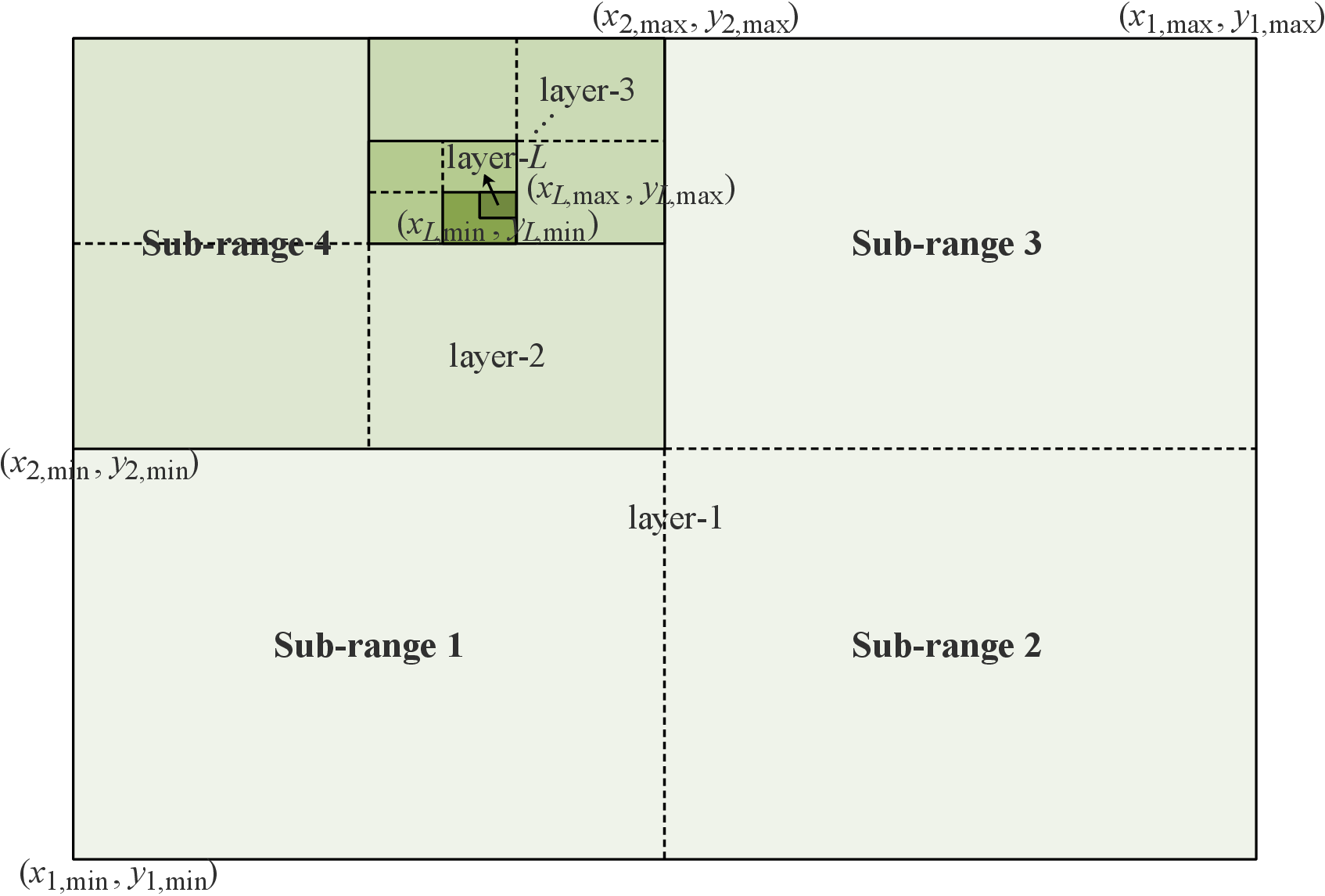}
\caption{Illustration for the proposed hierarchical beam training procedure.}
\label{NFCBeamTraining}
\end{center}
\end{figure}

As illustrated in Fig. {\ref{NFCBeamTraining}}, we divide the sampling range parallel to the XY-plane into four sub-ranges in each layer, and find the range corresponding to the optimal codeword as the sampling range of the next layer.
More particularly, in the $l$-th layer, the rectangular sampling range defined by the coordinate pairs $\left( x^k_{\min}, y^k_{\min} \right)$ and $\left( x^k_{\max}, y^k_{\max} \right)$, $k \in \left\{ {\rm{B}}, {\rm{U}} \right\}$, is divided into the following four sub-ranges:
\begin{subequations}\label{SubRange}
\begin{align}
1) \; & \left( x^k_{\min}, y^k_{\min} \right), \; \left( x^k_{\min} + x_{\Delta}, y^k_{\min} + y_{\Delta}  \right), \\
2) \; & \left( x^k_{\min} + x_{\Delta} , y^k_{\min} \right),  \; \left( x^k_{\max}, y^k_{\min} + y_{\Delta}  \right), \\
3) \; & \left( x^k_{\min} + x_{\Delta} , y^k_{\min} + y_{\Delta}  \right), \; \left( x^k_{\max}, y^k_{\max}  \right), \\
4) \; & \left( x^k_{\min} , y^k_{\min} + y_{\Delta} \right),  \; \left( x^k_{\min} + x_{\Delta} , y^k_{\max}  \right),
\end{align}
\end{subequations}
where $x_{\Delta} = \frac{1}{2}\left[x^k_{\max}- x^k_{\min} \right]$ and $y_{\Delta} = \frac{1}{2}\left[y^k_{\max}- y^k_{\min} \right]$.
Therefore, the sub-codebook for sub-range $i$ of ${\rm{B}}$ and sub-range $j$ of ${\rm{U}}$ in the $l$-th layer, denoted by ${\cal{F}}_{\rm{NN}} \left( l, i, j \right), i, j =1,2,3,4$, can be generated based on ({\ref{Codebook_NN}}).
Each column in ${\cal{F}}_{\rm{NN}} \left( l, i, j \right)$ corresponds to a codeword ${\bm{\phi }}_{l, i, j, s}$, with $s$ denoting the column index in sub-codebook ${\cal{F}}_{\rm{NN}} \left( l, i, j \right)$. 
Based on the given beamforming matrix ${\bf{W}}$ and codeword ${\bm{\phi }}_{l, i, j, s}$,  the achievable rate can be calculated as 
\begin{equation} R \left( {\bf{W}}, {\rm{diag}}\left( {\bm{\phi }}_{l, i, j, s} \right) \right). \end{equation} 

By adopting the ES method, the optimal codeword can be obtained for the first layer, as well as the optimal sampling sub-range pair $ \left \langle i, j \right \rangle$, which will be the sampling range in the next layer.
This process is repeated until the maximum number of layers $L$ is reached.

\begin{algorithm}[!t]
\caption{ Hierarchical Beam Training Scheme for NN }
\label{alg:2}
\begin{algorithmic}[1]
\REQUIRE sampling range $x(y)^{k}_{\min(\max)}$, $ k \in \left\{ {\rm{B}}, {\rm{U}}\right\} $, number of sampled points in each layer $S_x $ and $S_y $, maximum layer $L_{\max}$.
\ENSURE ${\bf{\Theta }}_{\rm{opt}}$, $R_{\rm{opt}}$.
\STATE {\bf Initialization:} $l=1$, $x(y)^{k}_{\min(\max)}\left( l \right) = x(y)^{k}_{\min(\max)}$, $ k \in \left\{ {\rm{B}}, {\rm{U}}\right\} $, $ R_{\rm{opt}} = 0 $.
\REPEAT
\STATE Divide the sampling ranges corresponding to both the BS and the user into four sub-ranges according to (\ref{SubRange});
\STATE Generate sub-codebooks ${\cal{F}}_{\rm{NN}} \left( l, i, j \right), i, j =1,2,3,4$, based on ({\ref{Codebook_NN}}), where $i$ is the sampling sub-range index for ${\rm{B}}$ and $j$ is the sampling sub-range index for ${\rm{U}}$;
\STATE Calculate $R \left( {\bf{W}}, {\bf{\Theta }} \right)$ for each codeword;
\STATE Find the maximum $R \left( {\bf{W}}, {\bf{\Theta }} \right)$, the corresponding codeword ${\bf{\Theta }}_{\rm{opt}}$, and the corresponding optimal sampling sub-range index pair $ \left \langle i, j \right \rangle$;
\STATE Update $x(y)^{\rm{B}}_{\min(\max)}\left( l+1 \right) = x(y)^{\rm{B}}_{\min(\max)} \left( l, i \right)$;
\STATE Update $x(y)^{\rm{U}}_{\min(\max)}\left( l+1 \right) = x(y)^{\rm{U}}_{\min(\max)} \left( l, j \right)$;
\STATE $l \leftarrow l + 1$; 
\UNTIL $l = L_{\max}$.
\end{algorithmic}
\end{algorithm}

The hierarchical beam training procedure is summarized in {\bf{Algorithm {\ref{alg:2}}}}.
We define the size of the codebook as the number of candidate codewords, i.e., the number of column vectors. 
There are 16 sub-codebooks in each layer, and the size of each sub-codebook is $\left( S_x  S_y \right)^2$.
Therefore, with a given maximum number of training layers $L$, the training overhead can be calculated as $16L\left( S_x  S_y \right)^2$.
For the same sampling grid resolution and hence beam training accuracy, the training overhead of ES will be $ 4^{L+1} \left( S_x  S_y \right)^2 $.

\subsubsection{ Two-Stage Beam Training Scheme for FN and NF }

The size of codebooks ${\cal{F}}_{\rm{NF}}$ and ${\cal{F}}_{\rm{FN}}$ is determined by the product of the number of RIS elements and the number of sample points, which can be extremely large especially for large RISs.
Thus, while ES is a straightforward way to find the optimal codeword from the predefined codebooks, the training overhead and consumed time can be unacceptable. 
We therefore propose a two-stage beam training scheme for the FN and NF channel models.
Specifically, all candidate angles are swept through in {\textit{stage one}}, and the distance is determined in {\textit{stage two}} based on the selected direction.
The details are given as follows.

{\bf{Stage 1 }} Angular-domain beam sweeping: 
In the first stage, we aim to find the optimal beamforming angle based on the far-field codebook given in (\ref{Codebook_FFC}), with the objective of maximizing the achievable rate in (\ref{DataRate}).
All sampled angles will be swept through in this stage, and the one yielding the maximum achievable rate will be selected. 
Therefore, the selected indices in the angular-domain sweeping stage can be expressed mathematically as
\begin{equation}
\begin{array}{l}
\;\;\;\; \left \langle m_x, m_y \right \rangle_{\rm{opt}}  \\
= {\rm{argmax}}\Big\{ {{\bf{b}}}\left( {\beta_{m_x}, \delta_{m_y} } \right) \in {\cal{F}}_{\rm{FF}} \left |  R \left( {\bf{W}}, {\bf{\Theta }}_{{m_x, m_y }} \right)   \right.\Big\},
\end{array}
\end{equation}
where ${\bf{\Theta }}_{{m_x, m_y }} = {\rm diag} \left( {{\bf{b}}}\left( {\beta_{m_x}, \delta_{m_y}} \right) \right)$. 
Denote the optimal index pair as $\left \langle m'_x, m'_y \right \rangle$. 

{\bf{Stage 2 }} Distance-domain sampling: 
In NFC, the beam can be designed to point at a particular location, rather than just a specific angle.
Thus, in the second stage, we aim to determine the optimal focusing point for the optimal direction.
With a given angle pair $\left \langle m'_x, m'_y \right \rangle$, the selected sampling point in the distance domain can be determined by
\begin{equation}
\begin{array}{l}
\;\;\;\; \left \langle  x_{s_x}^k, y_{s_y}^k \right \rangle _{\rm{opt}}  \\
= {\rm{argmax}}\left\{ \left(  x_{s_x}^k, y_{s_y}^k \right) \in {\cal{D}}_{k} \left |  R \left( {\bf{W}}, {\bf{\Theta }}_{x_{s_x}^k, y_{s_y}^k} \right)   \right.\right\},
\end{array}
\end{equation}
where ${\cal{D}}_{k}$ lies in the sampling range for $k$, $k \in \left\{ {\rm{B}}, {\rm{U}} \right\}$, and ${\bf{\Theta }}_{x_{s_x}^k, y_{s_y}^k} = {\rm diag} \left( {{\bf{c}}}\left(  x_{s_x}^k, y_{s_y}^k, \beta_{m'_x}, \delta_{m'_y} \right) \right)$.

Similar to the analysis in Section {\ref{HierarchicalTraining}}, we adopt a hierarchical beam training scheme in this stage to strike a balance between the training overhead and accuracy. 
Thus, the two-stage beam training procedure is summarized in {\bf{Algorithm {\ref{alg:1}}}}.
The overhead of the proposed two-stage beam training scheme is $M+4L{S_x}{S_y} $, while that of the ES method is $M \times 4L{S_x}{S_y}$.

\begin{myRemark} \label{Remark_MultiUser}
The proposed beam training algorithms can be easily extended to scenarios with multiple users by dividing the RIS into multiple sub-surfaces, where each sub-surface implements the proposed beam training schemes to realize beam alignment for each user.
\end{myRemark}

\begin{algorithm}[!t]
\caption{ Two-Stage Beam Training Scheme for Hybrid Cascaded Channel Models }
\label{alg:1}
\begin{algorithmic}[1]
\REQUIRE $M_x$, $M_y$, sampling range $x(y)^{k}_{\min(\max)}$, $ k \in \left\{ {\rm{B}}, {\rm{U}}\right\} $, number of sampled points in each layer $S_x $ and $S_y $, maximum layer $L_{\max}$.
\ENSURE ${\bf{\Theta }}_{\rm{opt}}$, $R_{\rm{opt}}$.
\STATE {\bf Initialization:} $m_x=1$, $m_y=1$, $ R_{\rm{opt}} = 0 $.
\STATE {\bf{Stage 1:}}
\STATE Generate sub-codebook ${\cal{F}}_{\rm{Hb}}^1$ based on ({\ref{Codebook_Hy1}});
\REPEAT
\REPEAT
\IF {$ R \left( {\bf{W}}, {\bf{\Theta }}_{{m_x, m_y }} \right) > R_{\rm{opt}} $}
\STATE $R_{\rm{opt}} = R \left( {\bf{W}}, {\bf{\Theta }}_{{m_x, m_y }} \right)$;
\STATE $\left \langle m_x, m_y \right \rangle_{\rm{opt}} = \left \langle { {m_x}, {m_y} } \right \rangle$;
\ENDIF
\STATE $m_x \leftarrow m_x + 1$;
\UNTIL $m_x = M_x$;
\STATE $m_y \leftarrow m_y + 1$.
\UNTIL $m_y = M_y$;
\STATE {\bf{Stage 2:}}
\REPEAT
\STATE Divide the sampling ranges corresponding to the BS (NF Case) or the user (FN Case) into four sub-ranges according to (\ref{SubRange});
\STATE Generate sub-codebooks ${\cal{F}}_{\rm{Hy}}^2 \left( l, i \right), i =1,2,3,4$, based on ({\ref{Codebook_Hy2}}), where $i$ is the sampling sub-range index for ${\rm{B}}$ (NF Case) or ${\rm{U}}$ (FN Case);
\STATE Generate ${\cal{F}}_{{\rm{NF}}/{\rm{FN}}} = {\left\{ {\cal{F}}_{\rm{Hb}}^1\right\}} _{\left \langle m_x, m_y \right \rangle_{\rm{opt}}} \star {\cal{F}}_{\rm{Hb}}^{2, k}$;
\STATE Calculate $R \left( {\bf{W}}, {\bf{\Theta }}_{x_{s_x}^k, y_{s_y}^k} \right)$ for each codeword;
\STATE Find the maximum $R \left( {\bf{W}}, {\bf{\Theta }}_{x_{s_x}^k, y_{s_y}^k} \right)$, the corresponding codeword ${\bf{\Theta }}_{\rm{opt}}$, and the corresponding optimal sampling sub-range index $ i $;
\STATE Update $x(y)^{k}_{\min(\max)}\left( l+1 \right) = x(y)^{k}_{\min(\max)} \left( l, i \right)$;
\STATE $l \leftarrow l + 1$; 
\UNTIL $l = L_{\max}$.
\end{algorithmic}
\end{algorithm}

\subsection{Closed-form Solution for $\bf{U}$}

With the estimated channel information acquired through the beam training, we can perform the BS beamforming and the user combining design.
Although the combining matrix $\bf{U}$ is not present in the objective function (\ref{DataRate}), it will have an impact on the BS beamforming design and the achievable rate.
Therefore, we derive the optimal combining matrix $\bf{U}$ at the user below.
The principle of combining matrix design is to recover the original transmitted signals, which can be transformed into a mean square error (MSE) minimization problem.
The MSE matrix of the user is given by
\begin{equation}\label{MSE}
\begin{array}{l}
{\bf{E}} = {\mathbb{E}}\left[  \left( {\bf{y}} - {\bf{x}} \right)  \left({\bf{y}} - {\bf{x}} \right)^H\right] \\
\;\; = \left( {\bf{U}}^H {\bf{H}} {\bf{W}} - {\bf{I}}_{q} \right) \left( {\bf{U}}^H {\bf{H}} {\bf{W}} - {\bf{I}}_{q} \right) ^H + \sigma^{2}{\bf{U}}^H{\bf{U}},
\end{array}
\end{equation}
and thus the mathematical form of the combining matrix optimization problem can be written as
\begin{equation}\label{P_U}
({\rm{P1}}) \; \min\limits_{\bf{U}} {\rm{Tr}}\left( {\bf{E}} \right).
\end{equation}

With given $\bf{W}$ and ${\bf{\Theta }}$, the optimal solution ${\bf{U}}$ of problem (\ref{P_U}) can be obtained by solving ${\partial {\rm{Tr}}\left( {\bf{E}} \right)}/{\partial {\bf{U}}} = {\bf{0}}$.
After some simple derivations, the optimal combining matrix ${\bf{U}}$ is found to be
\begin{equation}\label{Opt_U}
{\bf{U}}_{\rm{opt}} = \left( {\bf{H}} {\bf{W}} {\bf{W}}^H {\bf{H}}^H + \sigma^2{\bf{I}}_{N_{\rm{U}}}\right)^{-1}{\bf{H}} {\bf{W}},
\end{equation}
which is necessary for the optimal value of ${\bf{W}}$ to achieve the maximum of (\ref{DataRate}).

\subsection{Active Beamforming Matrix Optimization }

To solve the transmit beamforming optimization problem, we introduce an auxiliary matrix ${\bf{F}} \succeq 0$, and reformulate the original optimization as follows by exploiting the relationship between the achievable rate and the MSE {\cite{TSP.2011.2147784}}
\begin{equation}\label{P_Re}
\begin{array}{l}
({\rm{P'}})\; \max \limits_{ {\bf{F}}, \bf{W}, {\bm{\phi }} } f \left( {\bf{F}}, {\bf{W}}, {\bm{\phi }} \right),  \\
s.t. \;\; (\ref{P_max}), (\ref{CodebookConstraint}),
\end{array}
\end{equation}
where $f \left( {\bf{F}}, {\bf{W}}, {\bm{\phi }} \right)$ is given by
\begin{equation}\label{f}
f \left( {\bf{F}}, {\bf{W}}, {\bm{\phi }} \right) = \log\left|{\bf{F}} \right| - {\rm{Tr}}\left( {\bf{F}}{\bf{E}} \right).
\end{equation}

The optimal ${\bf{F}}$ of problem (\ref{P_Re}) can be obtained by solving ${\partial f}/{\partial {\bf{F}}} = {\bf{0}}$, which is given by ${\bf{F}} = {\bf{E}}^{-1}$.
With given ${\bf{F}}$, ${\bf{U}}$ and ${\bm{\phi }}$, the beamforming optimization problem can be expressed by substituting ${\bf{E}}$ into (\ref{f}) and removing the constant terms:
\begin{equation}\label{P_W}
\begin{array}{l}
({\rm{P2}})\;  \min \limits_{ \bf{W} }   {\rm{Tr}}\left( {\bf{W}}^H {\bf{H}}^H {\bf{U}} {\bf{F}} {\bf{U}}^H {\bf{H}} {\bf{W}}  \right) -2  {\rm{Tr}}\left(\Re \left( {\bf{F}} {\bf{U}}^H {\bf{H}} {\bf{W}} \right) \right)   \\
s.t. \;\; (\ref{P_max}),
\end{array}
\end{equation}
which is a convex optimization problem and its optimal solution can be obtained using standard methods.

\subsection{Overall Algorithm}  

\begin{figure}[t]
\begin{center}
\includegraphics[width=8.5cm]{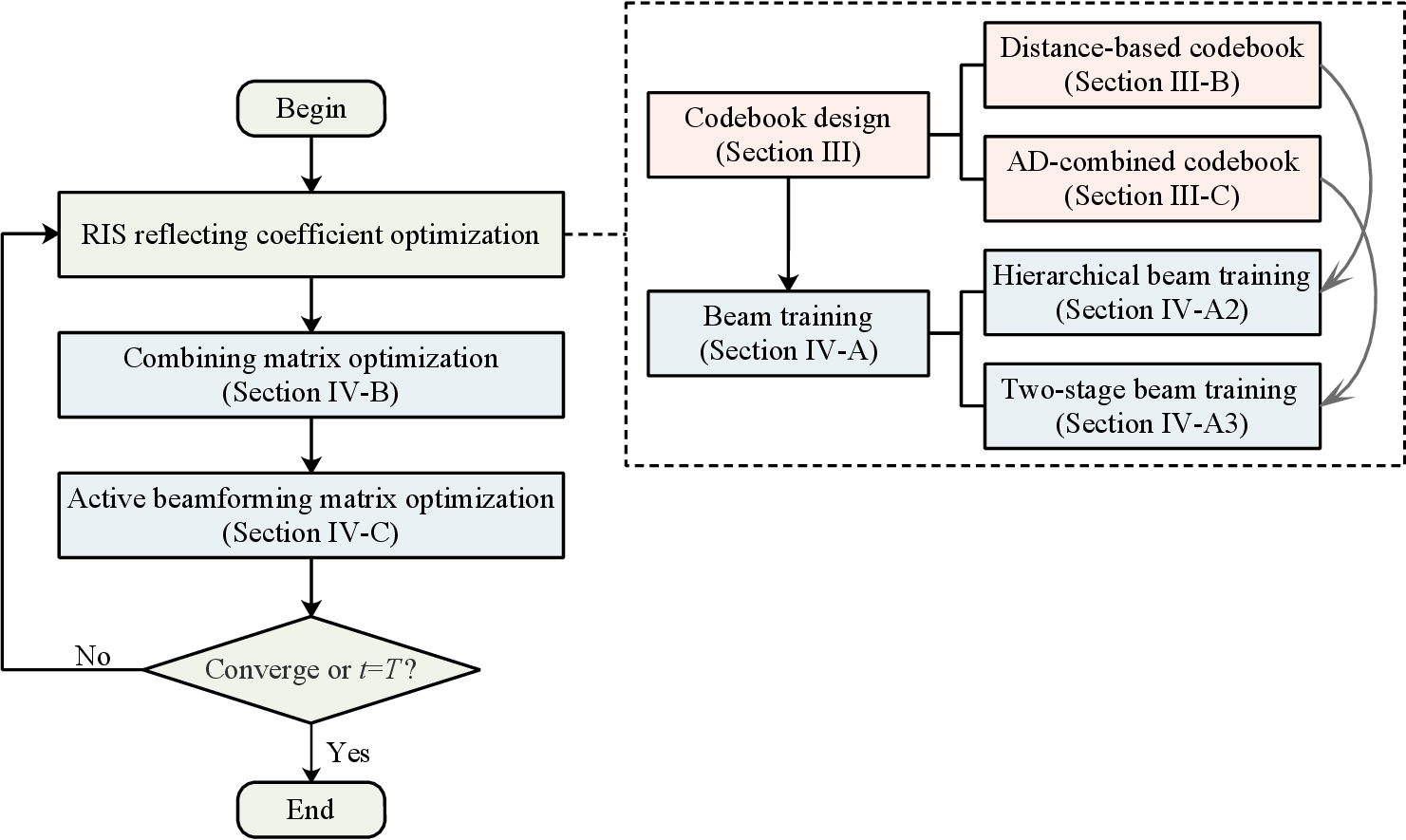}
\caption{Illustration for the proposed algorithm.}
\label{Fig_AlgorithmChart}
\end{center}
\end{figure}

The overall alternating optimization algorithm for maximizing the achievable rate is summarized in {\bf Algorithm {\ref{alg:3}}}, where the optimal solutions in each iteration are the inputs for the next iteration.
First, with given ${{\bf{U}}^{\left( t \right)}}$ and ${{\bf{\Theta}}^{\left( t \right)}}$ at the $t$-th iteration, the optimal ${{\bf{\Theta}}^{\left( {t + 1} \right)}}$ can be obtained by carrying out the beam training scheme based on {\bf Algorithm {\ref{alg:1}}} or {\bf Algorithm {\ref{alg:2}}}.
The estimated channel information can also be obtained in this step.
Then, with given ${{\bf{\Theta}}^{\left( t+1 \right)}}$ and ${{\bf{W}}^{\left( t \right)}}$, we calculate and update the combining matrix ${{\bf{U}}^{\left( {t + 1} \right)}}$ according to (\ref{Opt_U}).
Although ${\bf{U}}$ is not included in the expression of the objective function in (\ref{DataRate}), it will serve as input for the next step in solving ${{\bf{W}}}$.
Last, with given ${{\bf{U}}^{\left( {t + 1} \right)}}$, ${{\bf{\Theta}}^{\left( t+1 \right)}}$ and ${{\bf{W}}^{\left( t \right)}}$, we first obtain the reformulated objective function by substituting $ {\bf{E}}^{\left( {t } \right)} $ in (\ref{MSE}) into (\ref{f}), and then obtain the optimized beamforming matrix ${{\bf{W}}^{\left( {t + 1} \right)}}$ by solving problem (\ref{P_W}).
This process is repeated until a maximum number of iterations is reached or the convergence condition is satisfied. 
The proposed algorithm is summarized in Fig. {\ref{Fig_AlgorithmChart}}.

The computational complexity of the beam training mainly comes from sorting, which requires ${\cal O}\left( M \right)$ operations for angular-domain beam sweeping, ${\cal O}\left( 16L\left( S_x  S_y \right)^2 \right)$ operations for the hierarchical beam training scheme, and ${\cal O}\left( M+4L{S_x}{S_y} \right)$ operations for the two-stage beam training algorithm.
Solving problem ${\left( {\rm{P1}} \right)}$ for the combining matrix involves matrix inversion, leading to a computational complexity of ${\cal O}\left( N_{\rm{U}}^3\right)$ {\cite{TWC2021_Double_RIS }}.
According to {\cite{LOBO1998193}}, the worst-case complexity for solving problem ${\left( {\rm{P2}} \right)}$ is ${\cal O}\left( q^3 N_{\rm{B}}^3 \right)$ using efficient interior-point methods. 
Thus, the overall computational complexity of the proposed algorithm is ${\cal O}\left( I_3\left( M + q^3 N_{\rm{B}}^3 + N_{\rm{U}}^3 \right) \right)$ for FF with angular-domain beam sweeping, ${\cal O}\left( I_3\left( 16L\left( S_x  S_y + q^3 N_{\rm{B}}^3 + N_{\rm{U}}^3 \right) \right) \right)$ for NN with hierarchical beam training, and ${\cal O}\left( I_3\left( M+4L{S_x}{S_y} + q^3 N_{\rm{B}}^3 + N_{\rm{U}}^3 \right) \right)$ for NF and FN with two-stage beam training, with $I_3$ representing the number of iterations required for the convergence of {\bf Algorithm {\ref{alg:3}}}.

\begin{algorithm}[!t]
\caption{ Alternating Optimization-based Algorithm for Maximizing Achievable Rate }
\label{alg:3}
\begin{algorithmic}[1]
\REQUIRE $R \left( {\bf{W}}, {\bf{\Theta }} \right)$, maximum iteration times ${T}$, convergence threshold ${\zeta}$.
\ENSURE ${{\bf{U}}_{\rm{opt}}}$, ${{\bf{\Theta}}_{\rm{opt}}}$, ${{\bf{W}}_{\rm{opt}}}$.
\STATE {\bf Initialization:} $ t = 0 $, ${{\bf{U}}^{\left( t \right)}}$, ${{\bf{\Theta}}^{\left( t \right)}}$, ${{\bf{W}}^{\left( t \right)}}$, $R^{\left( {t } \right)} \left( {\bf{W}}, {\bf{\Theta }} \right)$.
\WHILE { $t \le T$ and $\Gamma \ge \zeta$  }
\STATE  For given ${{\bf{U}}^{\left( {t } \right)}}$ and ${{\bf{W}}^{\left( {t } \right)}}$, perform beam training based on {\bf Algorithm {\ref{alg:1}}} or {\bf Algorithm {\ref{alg:2}}}, and obtain ${{\bf{\Theta}}^{\left( {t + 1} \right)}}$;
\STATE  For given ${{\bf{\Theta}}^{\left( t+1 \right)}}$ and ${{\bf{W}}^{\left( t \right)}}$, calculate and update ${{\bf{U}}^{\left( {t + 1} \right)}}$ according to (\ref{Opt_U});
\STATE  For given ${{\bf{U}}^{\left( {t + 1} \right)}}$, ${{\bf{\Theta}}^{\left( t+1 \right)}}$ and ${{\bf{W}}^{\left( t \right)}}$, update ${{\bf{W}}^{\left( {t + 1} \right)}}$ by solving problem (\ref{P_W});
\STATE Calculate ${ R \left( {\bf{W}}^{\left( {t + 1} \right)}, {\bf{\Theta }}^{\left( {t + 1} \right)} \right) }$;
\STATE Calculate $ \Gamma  = \frac{ {\left| R \left( {\bf{W}}^{\left( {t + 1} \right)}, {\bf{\Theta }}^{\left( {t + 1} \right)} \right) - R \left( {\bf{W}}^{\left( {t} \right)}, {\bf{\Theta }}^{\left( {t} \right)} \right) \right|} } {{\left| R \left( {\bf{W}}^{\left( {t } \right)}, {\bf{\Theta }}^{\left( {t } \right)} \right) \right|}} $;
\STATE  $t \leftarrow t + 1$;
\ENDWHILE
\end{algorithmic}
\end{algorithm}

\section{Numerical Results}

Numerical results are provided in this section to validate the effectiveness of our proposed approach.
The simulation configurations listed in {\bf{TABLE} {\ref{tab3}}} are used unless stated otherwise. 
The following parameter settings were used when executing the beam training schemes for both the hierarchical beam training algorithm for the NN channel model and the second beam training stage for hybrid cascaded channel models.
For hierarchical beam training, the sampled range in each direction is set as $\left[ -1000\lambda_c, 1000\lambda_c \right]$. 
 For example, denote the coordinate of the BS antenna midpoint as $\left( x_{\rm{BS}}, y_{\rm{BS}}, z_{\rm{BS}} \right)$, where $z_{\rm{BS}}$ is fixed.
The sampled ranges of the BS in the $x$- and $y$-directions are $\left[ x_{\rm{BS}}-1000\lambda_c, x_{\rm{BS}}+1000\lambda_c \right]$ and $\left[ y_{\rm{BS}}-1000\lambda_c, y_{\rm{BS}}+1000\lambda_c \right]$, respectively.
The same applies to the sampled ranges of the user.
The numbers of sampled points for hierarchical beam training in the $x$- and $y$-directions are both set as $S_x = S_y = 2$.
The simulation results are obtained by averaging over 500 channel realizations.

\begin{table}[!t]
  \caption{Simulation Configurations}
  \begin{center}
  \begin{tabular}{|c|c|}%
  \hline
   {\bf{Parameter}} & {\bf{Value}} \\
  \hline
  Carrier frequency & 30Ghz \\
  \hline
  Noise power & -105dBm \\
  \hline
  Power budget at the BS & 30 dBm \\
  \hline
  BS's coordinate & $\left( 0, 0, 0 \right)$m \\
  \hline
  User's coordinate & $\left( 24, 0, 0 \right)$m \\
  \hline
  RIS's coordinate & $\left( 10, 0, 8 \right)$m \\
  \hline
  Number of BS antennas & 16  \\ 
  \hline
  Number of user antennas & 8  \\ 
  \hline
  Number of RIS elements along $x$/$y$-axis & 60, 2  \\ 
  \hline
  Antenna spacing & $\lambda_c/2$ \\
  \hline
  Number of NLoS channels for far-field region & 2 \\
  \hline  
  \end{tabular}
  \label{tab3}
  \end{center}
\end{table}

\begin{figure}[t]
\centering
\includegraphics[width=7.0cm]{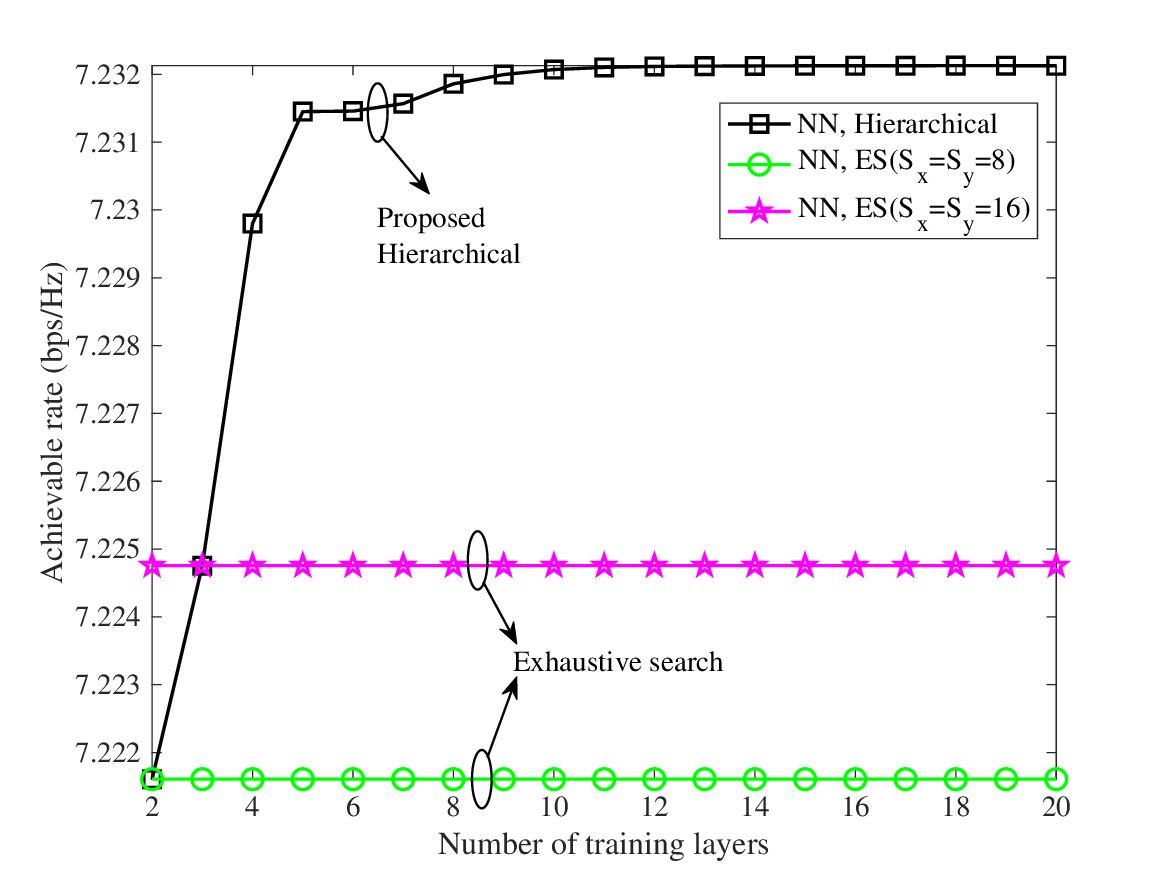}
\caption{Achievable rate vs. training layers of hierarchical beam training scheme.}
\label{Fig_TrainingLayer}
\end{figure}

To determine the number of training layers of the hierarchical beam training scheme, we show the relationship between the achievable rate and the number of training layers in Fig. {\ref{Fig_TrainingLayer}}.
It can be seen that when $L_{\max} \ge 12$, the achievable rate no longer increases with $L_{\max}$.
Although the achievable rate only improves slightly from $L_{\max} = 2$ to $L_{\max} = 12$, we set the total number of training layers hereafter as $L_{\max}=12$ for better performance.
Also, we depict the achievable rate obtained by the ES approach when the number of sampled points is set as 8 and 16 in each direction.
The size of the NFC codebook is $\left( S_x  S_y \right)^2$, so it can be extremely large for large $S_x$ and  $S_y$.
Thus, we only considered two scenarios with small $S_x$ and  $S_y$ to compare the performance between the proposed hierarchical beam training and ES approaches.
Note that we divide the whole sampling range into two sub-ranges in each direction, so there are 4 sampled points in each direction when $S_x = S_y = 2$ for hierarchical beam training.
Therefore, the proposed hierarchical algorithm achieves the same accuracy as ES with $S_x = S_y = 8$ when the number of training layers is 2, and it is also the same for $S_x = S_y = 16$ when the number of training layers is 3, which verifies the effectiveness of our proposed approach which has significantly reduced training costs.

\begin{figure}[t]
\centering
\includegraphics[width=7.0cm]{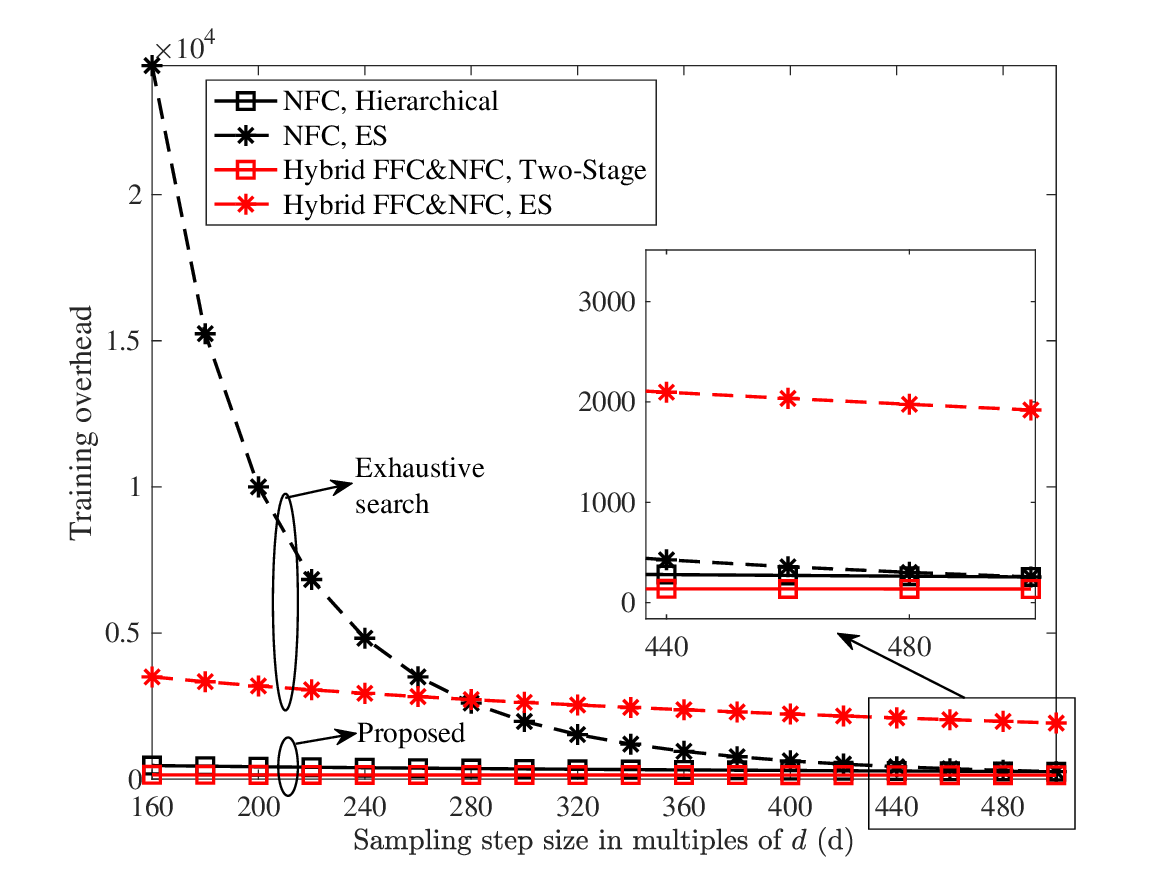}
\caption{Beam training overhead vs. sampling step size in multiples of $d$.}
\label{Fig_TrainingOverhead}
\end{figure}

The required beam training overhead of various training algorithms for different sampling step sizes is compared in Fig. {\ref{Fig_TrainingOverhead}}.
For NN channel modeling, the proposed hierarchical near-field beam training approach greatly reduces the training overhead compared to ES beam training, especially when the sampling step size is small.
When the step size is 160$d$, the overhead of the hierarchical beam training approach is only about 2\% of that required for ES. 
When the step size is 500$d$, these two schemes have the same training overhead, and the required number of training layers for hierarchical beam training is 1.
For hybrid FFC \& NFC channel models, hierarchical beam training always requires less overhead than ES, regardless of the sampling step size.

\begin{figure}[t]
\centering
\includegraphics[width=7.0cm]{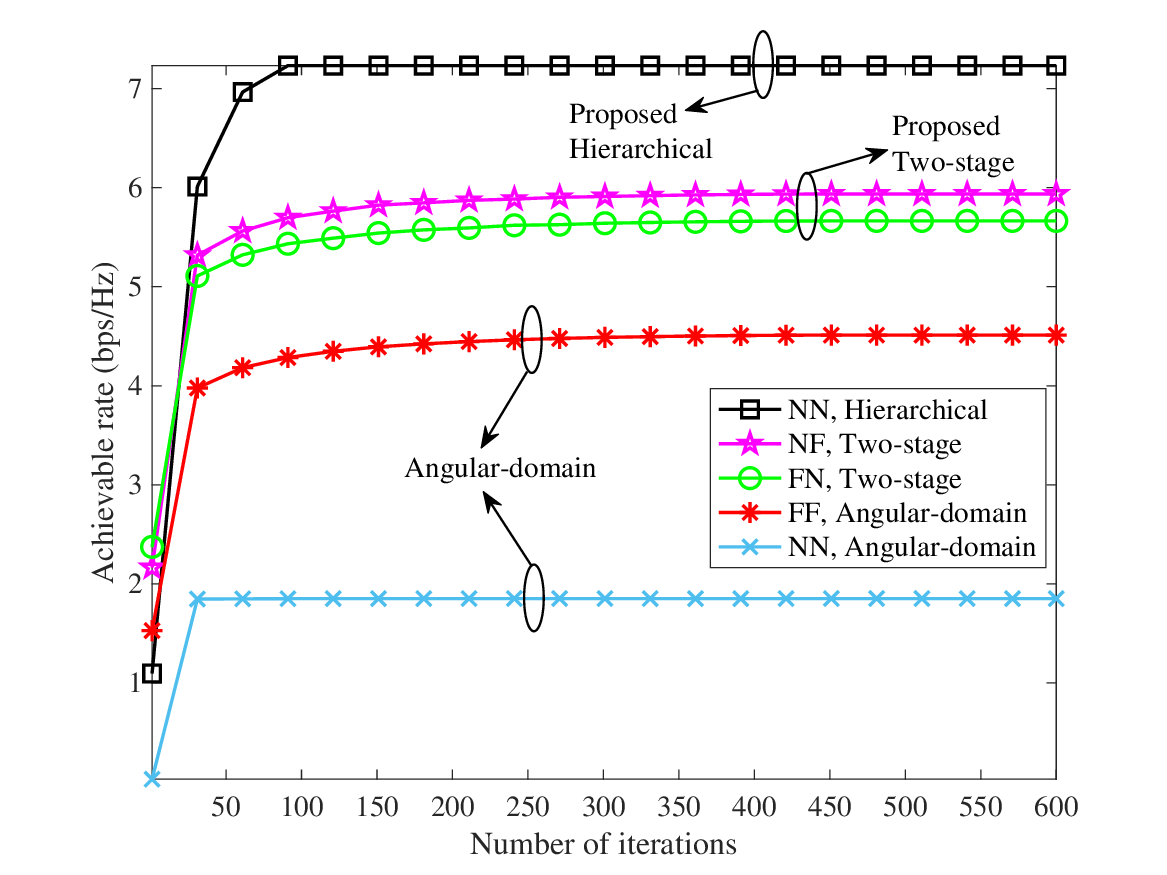}
\caption{Convergence performance of the proposed algorithm.}
\label{Fig_iteration}
\end{figure}

The convergence performance of the proposed AO-based algorithm is shown in Fig. {\ref{Fig_iteration}}.
Based on the simulation parameters provided earlier, we can calculate that both the BS-RIS and RIS-user links are within the near-field region. 
Since the NN channel model with hierarchical beam training is more accurate since it considers both angular and distance information of the cascaded links, it will provide a higher achievable rate.
Even though the NN channel model is adopted, the performance obtained with angular-domain beam training is poor, because distance information is not considered.
The FF channel model has the worst performance due to the accumulative channel estimation error of links on both sides of the RIS.
Furthermore, it can be seen that the achievable rate for the NF channel model is higher than that for FN, which is due to the fact that: 1) the BS and RIS are closer together than the RIS and user, and 2) the BS has more antennas than the user.
Thus, it is more critical for the BS-RIS link to adopt the near-field channel model to realize phase alignment.

\begin{figure}[t]
\centering
\includegraphics[width=7.0cm]{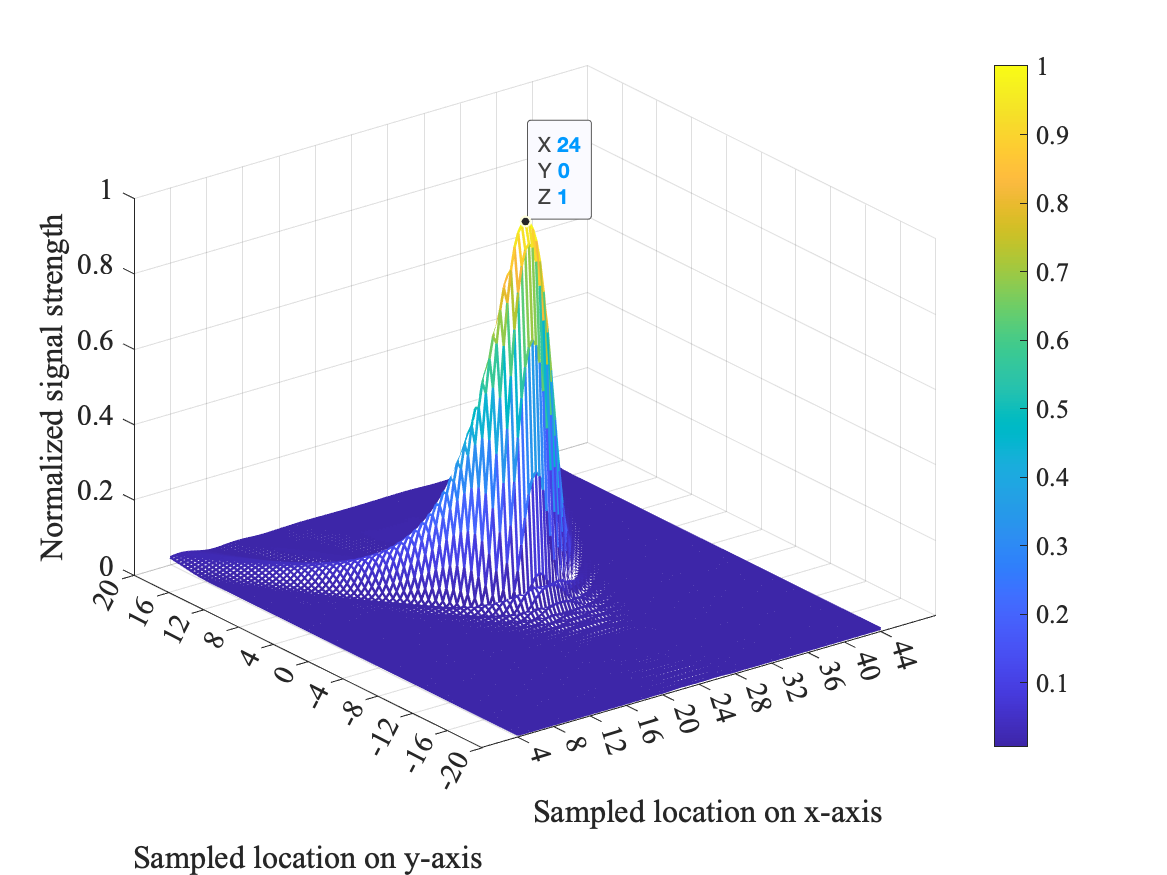}
\caption{Normalized received signal strength.}
\label{Fig_SignalStrength}
\end{figure}

For the NN channel model, the normalized received signal strength is depicted in Fig. {\ref{Fig_SignalStrength}}.
We can see that the maximum received power appears at coordinate (24, 0), which is exactly where the user is located.
Since the user's antennas are distributed parallel to the $x$-axis, the received signal strength is also dispersed along the $x$-axis.
Moreover, the RIS reflecting elements are more distributed along the $x$-axis direction, which will also lead to such a distribution of reflected signals.
Unlike FFC, NFC can achieve beam focusing, concentrating the energy of the beam at a specific location.

\begin{figure}[t]
\centering
\includegraphics[width=7.0cm]{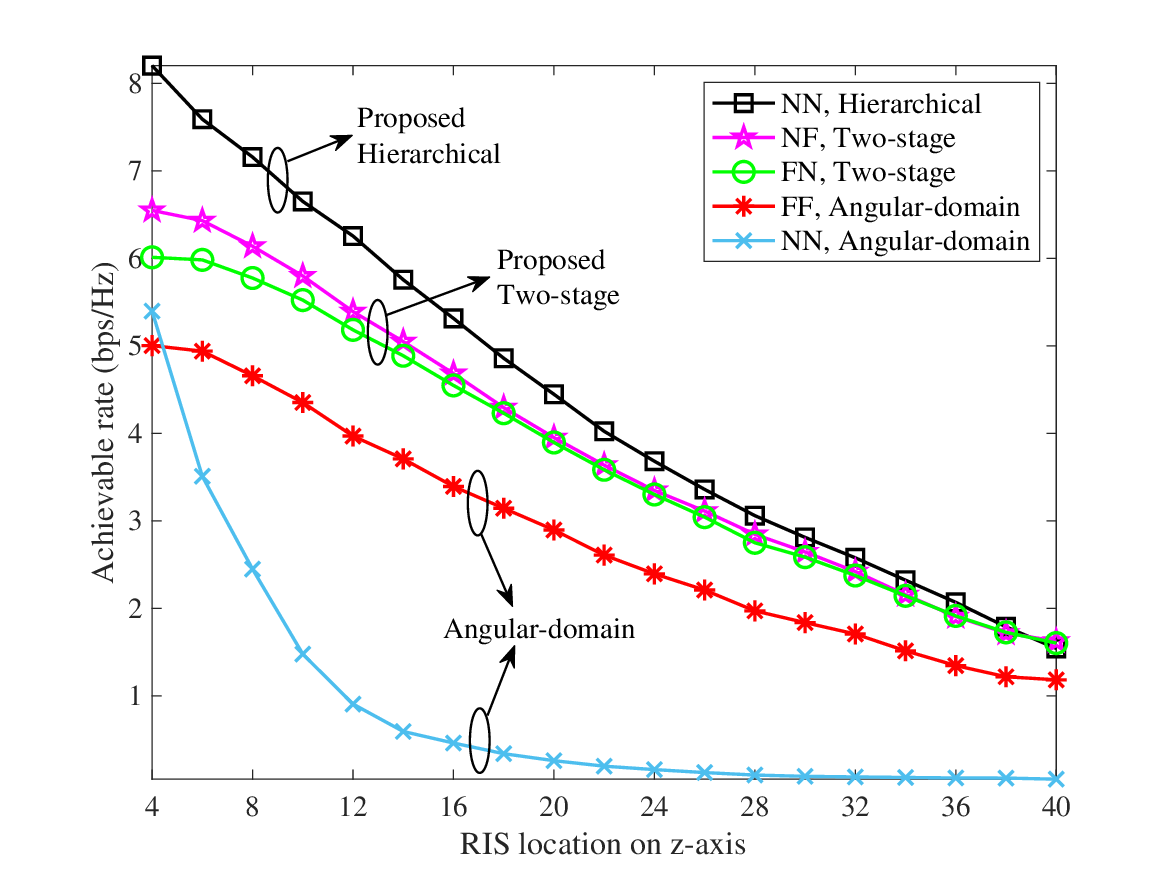}
\caption{Achievable rate vs. RIS location on $z$-axis.}
\label{Fig_RISLocation_z}
\end{figure}

Fig. {\ref{Fig_RISLocation_z}} shows the achievable rate as the RIS is located at different points in the $z$-direction.
Note that the $z$ coordinates of both the BS and the user are 0.
For smaller $z$ values, the closer the RIS is to the BS and user, the more obvious the near-field effect.
Therefore, the performance gain of adopting NN channel models and hierarchical beam training is more significant with smaller $z$, and the results are consistent with that in Fig. {\ref{Fig_iteration}} when $z=8$.
Moreover, as the distance between the RIS and BS and the user gradually increases, the performance gap due to different channel models and beam training methods gradually decreases, which also verifies that the near-field channel can be approximated as the far-field channel when the distance is large enough.

\begin{figure}[t]
\centering
\includegraphics[width=7.0cm]{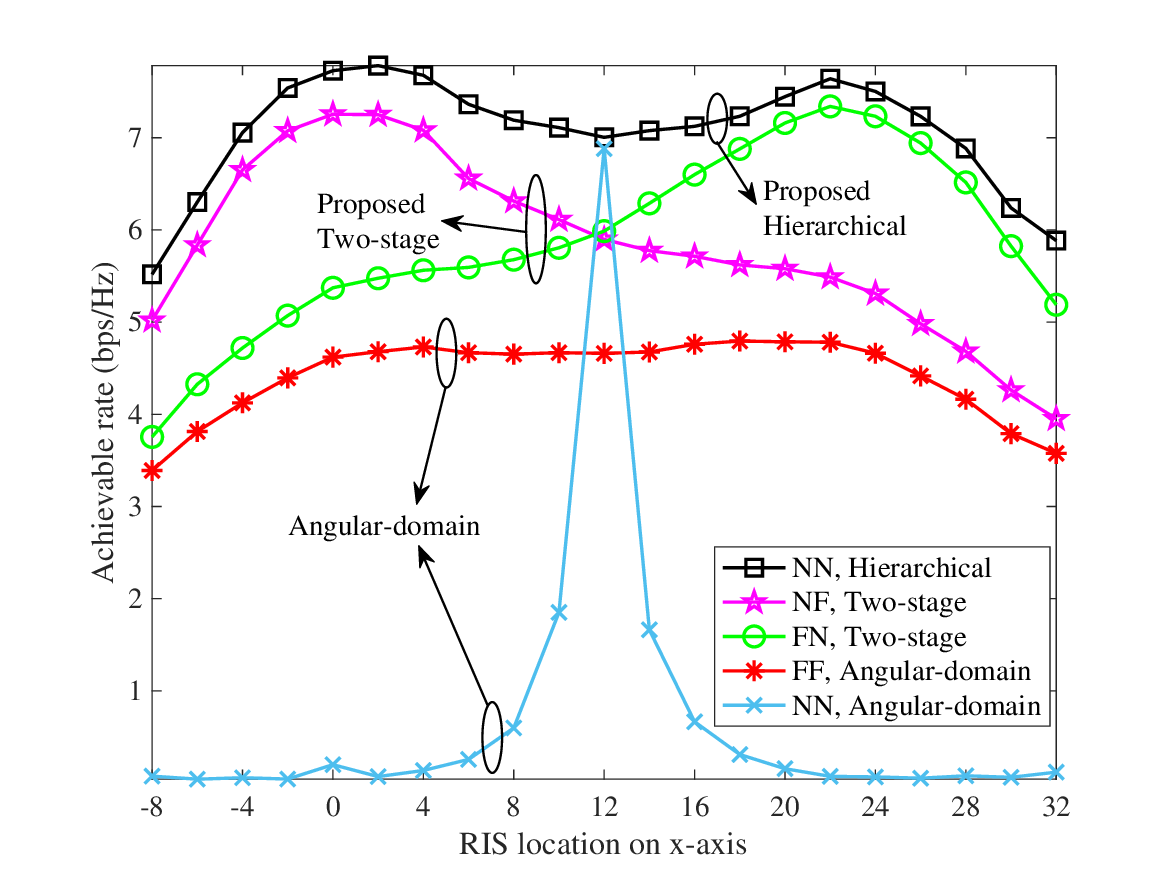}
\caption{Achievable rate vs. RIS location on $x$-axis.}
\label{Fig_RISLocation_x}
\end{figure}

The achievable rate obtained for different RIS locations along the $x$-direction is depicted in Fig. {\ref{Fig_RISLocation_x}}. 
It is not difficult to see that the NN channel model with hierarchical beam training always has the highest rate, regardless of the RIS location.
In addition, for the NN channel model, a peak appears on both the BS and user side, which is consistent with the conclusions for FFC.
Due to the larger number of antennas at the BS compared to the user, better transmission performance is obtained by deploying the RIS on the side near the BS.
For the NF channel model, the maximum achievable rate is obtained with the RIS near the BS, due to the more obvious near-field effect of the BS-RIS link.
As the RIS moves away from the BS, the achievable rate decreases, although this degradation in rate slows near the user's location.
For the FN channel model, in contrast to NF, the peak appears around the user.
When adopting the NN channel model and angular beam training, a peak occurs only when the RIS is deployed at an equal distance between the BS and the user, which means that angular beam training for the NN model is only useful in restricted scenarios.

\begin{figure}[t]
\centering
\includegraphics[width=7.0cm]{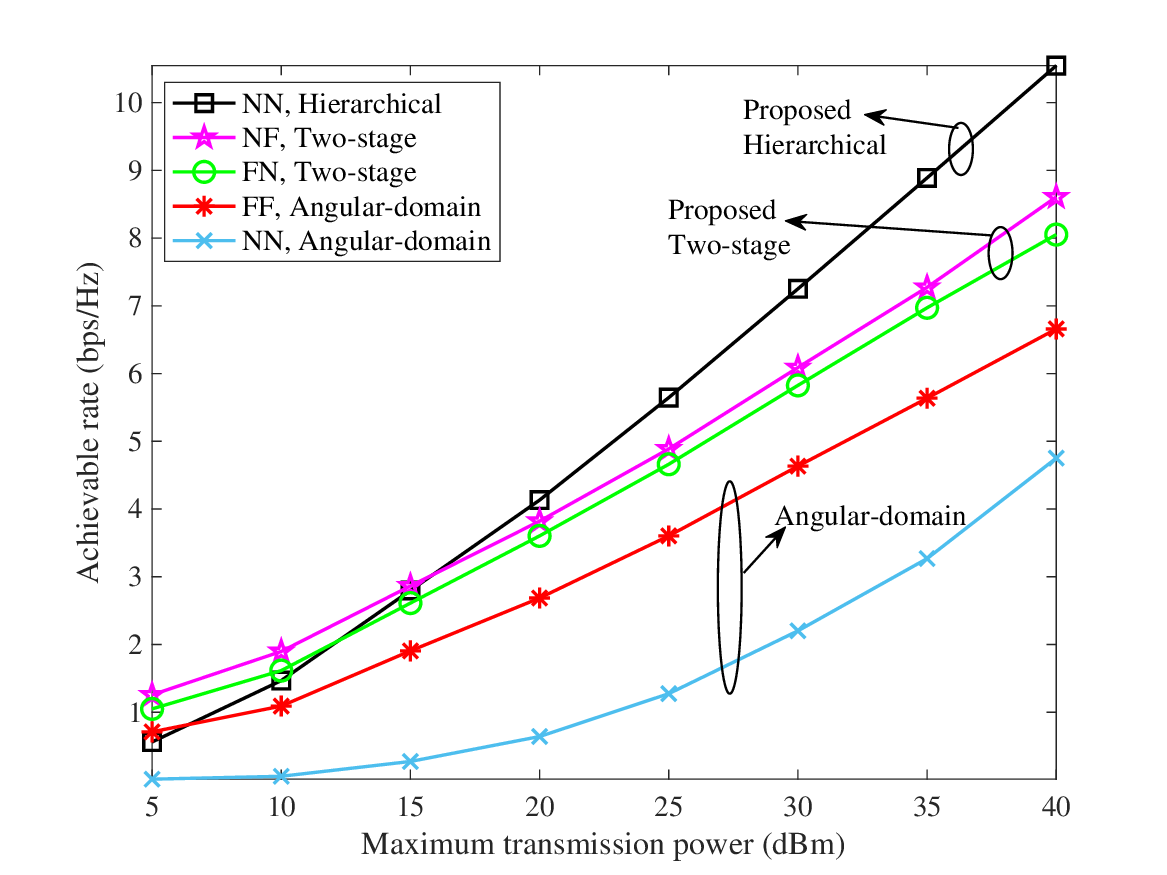}
\caption{Achievable rate vs. maximum transmission power at the BS.}
\label{Fig_Pmax}
\end{figure}

In Fig. {\ref{Fig_Pmax}}, we show the relationship between the achievable rate and the power budget at the BS.
As expected, regardless of the channel model or beam training approach,  the achievable rate increases with the BS power budget.
In addition, for low BS power,  NN with hierarchical beam training cannot fully leverage its advantages and has a lower achievable rate than NF and FN.
As the power increases, the achievable rate of NN increases the fastest, and its performance advantages become increasingly pronounced compared to the other models.
The rates achieved with the NF and FN channel models are nearly the same for all power levels, since the BS-RIS and RIS-user distances remain the same, and the same beam training methods are used for both.

\begin{figure}[t]
\centering
\includegraphics[width=7.0cm]{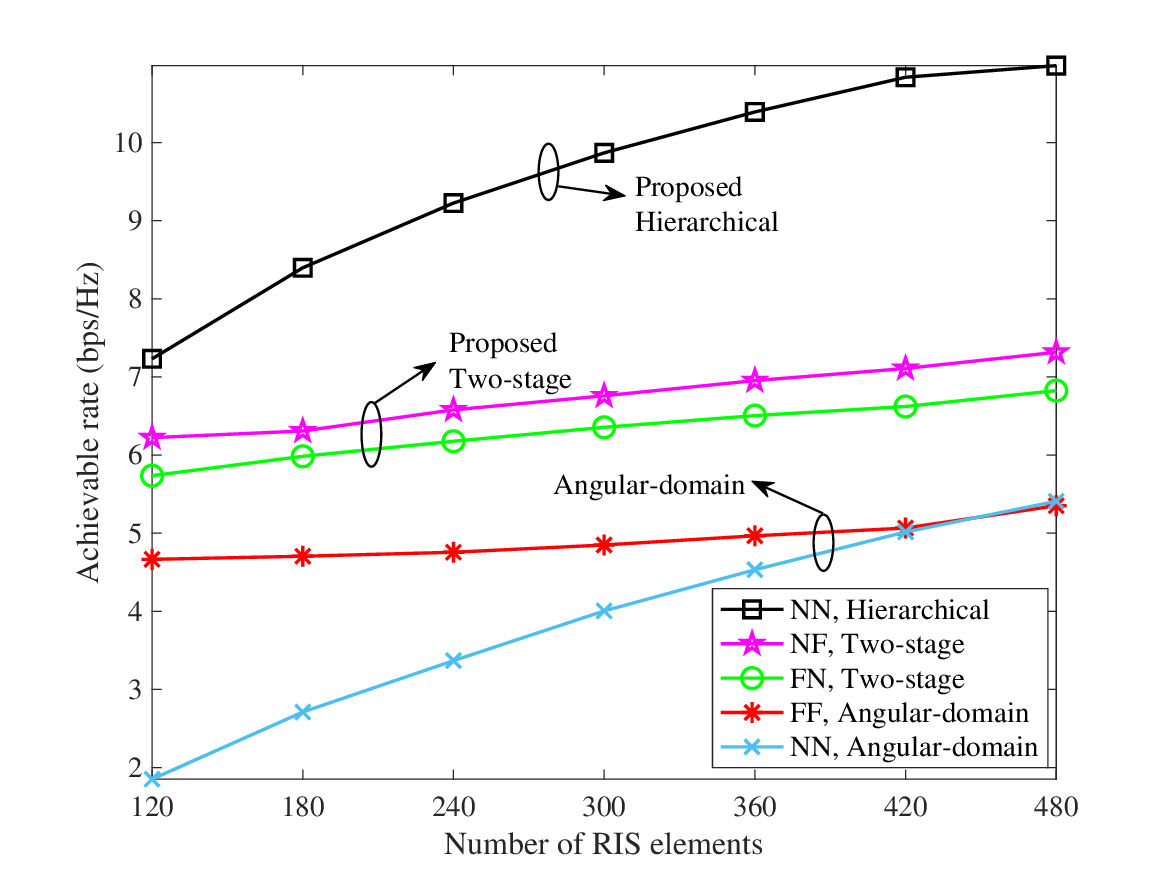}
\caption{Achievable rate vs. number of RIS elements.}
\label{Fig_RISelement}
\end{figure}

Fig. {\ref{Fig_RISelement}} shows the achievable rate for different numbers of RIS reflecting elements $M$. 
The achievable rate for all channel model and beam training approaches increases with $M$, due to the higher passive array gains that can be obtained from larger RIS.
The NN channel model achieves the fastest increase in rate as $M$ increases since the near-field effect is greater for larger RISs,  the NN model describes the channel more accurately, and the hierarchical beam training approach realizes a more effective beam focusing.
When the number of RIS elements is large enough, the achievable rate obtained using the NN channel model with angular-domain beam training will be superior to that of the FF model, even if the beam training approach is not specifically designed for the NN model.

\section{Conclusions}
In this article, we considered four different channel models for a RIS-assisted downlink MIMO system. 
According to the angular or distance information embedded in the received signals under these specific channel models, we designed different codebooks to match the beam steering vectors for the RIS. 
Based on the predesigned codebooks, we proposed two beam training approaches,  which were further used for RIS coefficient optimization. 
More specifically, for the NN channel model, we designed a distance-based codebook and proposed a hierarchical beam training algorithm to realize beam alignment while reducing the training overhead; for the NF and FN channel models, we designed a combined angular-distance codebook and proposed a two-stage beam training approach to separately realize beam alignment in the angular- and distance domains.
To maximize the achievable rate, we proposed an AO-based algorithm to carry out the multi-resource optimization in an iterative manner. 
Numerical results show that the proposed beam training approaches can obtain achievable rate performance similar to the ES method, while significantly reducing the training overhead. Our results demonstrate that the use of distance in addition to angular information can effectively improve the system performance with near-field channels.

\balance

\bibliographystyle{IEEEtran}
\bibliography{RIS-aided_Near-Field_MIMO_Communications}

\end{document}